# A high-throughput *ab initio* review of platinum-group alloy systems

Gus L.W. Hart,[1] Stefano Curtarolo,[2, *] Thaddeus B. Massalski,[3] and Ohad Levy[2]

[1]*Department of Physics and Astronomy, Brigham Young University, Provo UT 84602, USA*
[2]*Center for Materials Genomics and Department of Mechanical Engineering and Materials Science, Duke University, Durham NC 27708, USA*
[3]*Materials Science, Engineering and Physics, Carnegie Mellon University, Pittsburgh, PA 15213, USA*

We report a comprehensive study of the binary systems of the platinum group metals with the transition metals, using high-throughput first-principles calculations. These computations predict stability of new compounds in 37 binary systems where no compounds have been reported in the literature experimentally, and a few dozen of as yet unreported compounds in additional systems. Our calculations also identify stable structures at compound compositions that have been previously reported without detailed structural data and indicate that some experimentally reported compounds may actually be unstable at low temperatures. With these results we construct enhanced structure maps for the binary alloys of platinum group metals. These are much more complete, systematic and predictive than those based on empirical results alone.

## I. INTRODUCTION

The platinum group metals (PGMs), osmium, iridium, ruthenium, rhodium, platinum and palladium, are immensely important in numerous technologies, but the experimental and computational data on their binary alloys still contain many gaps. Interest in PGMs is driven by their essential role in a wide variety of industrial applications, which is at odds with their high cost. The primary application of PGMs is in catalysis, where they are core ingredients in the chemical, petroleum and automotive industries. They also extensively appear as alloying components in aeronautics and electronics applications. The use of platinum alloys in the jewelry industry also accounts for a sizeable fraction of its worldwide consumption, about 30% over the last decade [1]. The importance and high cost of PGMs motivate numerous efforts directed at more effective usage, or at the development of less-expensive alloy substitutes. Despite these efforts, there are still sizeable gaps in the knowledge about the basic properties of PGMs and their alloys; many of the possible alloy compositions have not yet been studied and there is a considerable difficulty in application of thermodynamic experiments because they often require high temperatures or pressures and very long equilibration processes.

The possibility of predicting the existence of ordered structures in alloy systems from their starting components is a major challenge of current materials research. Empirical methods use experimental data to construct structure maps and make predictions based on clustering of simple physical parameters. Their usefulness depends on the availability of reliable data over the entire parameter space of components and stoichiometries. Advances in first-principles methods for the calculation of materials properties open the possibility to complement the experimental data by computational results. Indeed many recent studies present such calculations of PGM alloy structures [2–25]. However, most of these studies consider a limited number of structures, at just a few stoichiometries of a single binary system or a few systems [3–17]. Some cluster expansion studies of specific binary systems include a larger set of structures, but limited to a single lattice type (usually, fcc) [18–23]. Realizing the potential of first-principles calculations to complement the lacking, or only partial, empirical data requires high-throughput computational screening of large sets of materials, with structures spanning all lattice types and including, in addition, a considerable number of off-lattice structures [2, 26–28]. Such large scale screenings can be used to construct low-temperatures binary phase diagrams. They provide insights into trends in alloy properties and indicate the possible existence of hitherto unobserved compounds [27]. A few previous studies implemented this approach to binary systems of specific metals, hafnium, rhenium, rhodium, ruthenium and technetium [24, 25, 29–31].

The capability to identify new phases is key to tuning the catalytic properties of PGM alloys and their utilization in new applications, or as reduced-cost or higher-activity substitutes in current applications. Even predicted phases that are difficult to access kinetically in the bulk may be exhibited in nanophase alloys [32] and could be used to increase the efficiency or the lifetimes of PGM catalysts. Given the potential payoff of uncovering such phases, we have undertaken a thorough examination of PGM binary phases with the transition metals, using the first-principles high-throughput (HT) framework AFLOW [33, 34]. We find new potentially stable PGM phases in many binary systems and, comparing experimental data with our predictions, we construct enhanced Pettifor-type maps that demonstrate new ordering trends and compound forming possibilities in these alloys.

## II. METHODS

Computations of the low-temperature stability of the PGM-transition metal systems were carried out using the HT framework AFLOW [33, 34]. For each of the 153 binary systems studied, we calculated the energies of more than 250 structures, including all the crystal structures reported for the system in the phase diagram literature [35, 36]

---

and additional structures from the AFLOWLIB database of prototypes and hypothetical hcp-, bcc- and fcc-derivative superstructures [33]. A complete list of structures examined for each binary system can be found on the on-line repository, www.aflowlib.org [34]. The low temperature phase diagram of a system is constructed as the minimum formation enthalpy convex hull from these candidate structures, identifying the ordering trends in each alloy system and indicating possible existence of previously unknown compounds. It should be noted that there is no guarantee that the true groundstates of a system will be found among the common experimentally observed structures or among small-unit-cell derivative structures. However, even if it is impossible to rule out the existence of additional unexpected groundstates, this protocol (searching many enumerated derivative structures [37] and exhaustively exploring experimentally reported ones) is expected to give a reasonable balance between high-throughput speed and scientific accuracy to determine miscibility (or lack thereof) in these alloys. In Ref. [2], it was shown that the probability of reproducing the correct ground state, if well defined and not ambiguous, is $\eta_C^* \sim 96.7\%$ ["reliability of the method," Eq. (3)].

The calculations of the structure energies were performed with the VASP software [38] with projector augmented waves pseudopotentials [39] and the exchange-correlation functionals parameterized by Perdew, Burke and Ernzerhof for the generalized gradient approximation [40]. The energies were calculated at zero temperature and pressure, with spin polarization and without zero-point motion or lattice vibrations. All crystal structures were fully relaxed (cell volume and shape and the basis atom coordinates inside the cell). Numerical convergence to about 1 meV/atom was ensured by a high energy cutoff (30% higher than the maximum cutoff of both potentials) and a 6000 **k**-point, or higher, Monkhorst-Pack mesh [41].

The presented work comprises 38,954 calculations, performed by using 1.82 million CPU/hours on 2013 Intel Xeon E5 cores at 2.2GHz. It was carried out by extending the pre-existing AFLOWLIB structure database [34] with additional calculations characterizing PGM alloys. Detailed information about all the examined structures can be found on the on-line repository, www.aflowlib.org [34], including input/output files, calculation parameters, geometry of the structures, energies and formation energies. In addition, the reader can prepare phase diagrams (as in figs. 5 to 12) linked to the appropriate structure URL locations.

The analysis of formation enthalpy is, by itself, insufficient to compare alloy stability at different concentrations and their resilience toward high-temperature disorder. The formation enthalpy, $\Delta H(A_x B_{1-x}) \equiv H(A_x B_{1-x}) - xH(A) - (1-x)H(B)$, represents the ordering-strength of a mixture $A_x B_{1-x}$ against decomposition into its pure constituents at the appropriate proportion $xA$ and $(1-x)B$ ($\Delta H$ is negative for compound forming systems). However, it does not contain information about its resilience against disorder, which is captured by the entropy of the system. To quantify this resilience we define the *entropic temperature*

$$T_s \equiv \max_i \left[ \frac{\Delta H(A_{x_i} B_{1-x_i})}{k_B \left[ x_i \log(x_i) + (1-x_i) \log(1-x_i) \right]} \right], \quad (1)$$

where $i$ counts all the stable compounds identified in the $AB$ binary system by the *ab initio* calculations, and the sign is chosen so that a positive temperature is needed for competing against compound stability. This definition assumes an ideal scenario [28] where the entropy is $S[\{x_i\}] = -k_B \sum_i x_i \log(x_i)$. This first approximation should be considered as indicative of a trend (see Fig. 1 of Ref. [28] and Fig. 1 below), which might be modified somewhat by a system specific thorough analysis of the disorder. $T_s$ is a concentration-maximized formation enthalpy weighted by the inverse of its entropic contribution. It represents the deviation of a system convex-hull from the purely entropic free-energy hull, $-TS(x)$, and hence the ability of its ordered phases to resist the deterioration into a temperature-driven, entropically-promoted, disordered binary mixture.

## III. HIGH-THROUGHPUT RESULTS

We examined the 153 binary systems containing a PGM and a transition metal, including the PGM-PGM pairs, (see Fig. 1). An exhaustive comparison of experimental and computational groundstates is given in tables I to VI. Convex hulls for systems which exhibit compounds are shown in the Appendix (figs. 5 to 12). These results uncover 37 alloy systems reported as non-compound forming in the experimental literature, but predicted computationally to have low-temperature stable compounds. Dozens of new compounds are also predicted in systems known to be compound forming.

The top panel of Fig. 1 gives a broad overview of the comparison of experiment and computation. Green circles (dark gray) indicate systems where experiment and computation agree that the system is compound forming. Light gray circles indicate agreement that the system is not compound-forming. The elements along the axes of this diagram are listed according to their Pettifor $\chi$ parameter [42, 43], leading, as expected, to compound-forming and non-compound forming systems separating rather cleanly into different broad regions of the diagram. Most of the compound-forming systems congregate in a large cluster on the left half of the diagram, and in a second smaller cluster at the lower right corner.

The systems for which computation predicts compounds but experiment does not report any are marked by red squares. As is clear in the top panel of Fig. 1, these systems, which harbor potential new phases, occur near the boundary between the compound-forming and non-compound-forming regions of the diagram. They also fill in several isolated spots where experiment reports no compounds in the compound-forming region (e.g., Pd-W, Ag-Pd), and bridge the gap between the large cluster of compound-forming systems, on the left side of the panel, and the small island of such systems at its center. The computations also predict ordered structures in most systems reported only with disordered phases (yellow circles in top panel of Fig. 1). Two disordered phases, $\sigma$ and $\chi$, turn up in the experimental literature on PGM alloys. In the HT search, we included all ordered realizations of these phases (the prototypes $Al_{12}Mg_{17}$ and $Re_{24}Ti_5$ are ordered versions of the $\chi$ phase and the $\sigma$ phase has 32 ordered realizations, denoted by $\sigma_{XXXXX}$ where $X = A, B$). In most of these systems we find one of these corresponding ordered structures to be stable. The only exception is the Cr-Ru system, where the lowest lying ordered phase is found just 4 meV/atom above the elements tie-line (yellow square in Fig. 1). These results thus identify the low temperature ordered compounds that underly the reported disordered phases. The calculated compound-forming regions are considerably more extensive than reported by the available experimental data, identifying potential new systems for materials engineering.

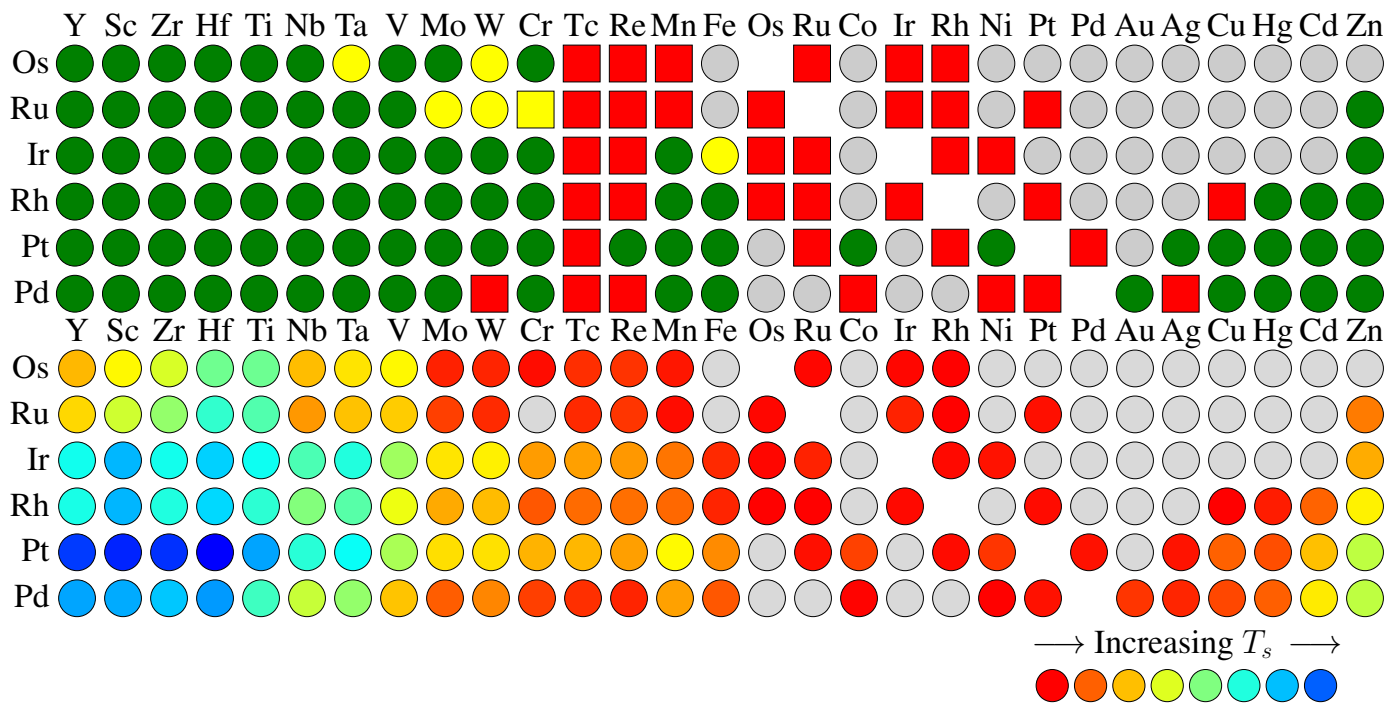

FIG. 1: Top panel: Compound-forming vs. non-compound-forming systems as determined by experiment and computation. Circles indicate agreement between experiment and computation, green for compound-forming systems, gray for non-compound forming systems. Yellow circles indicate systems reported in experiment to have disordered phases, for which low-energy compounds were found in this work. Ru-Cr is the only system (yellow square) experimentally reported to include a disordered phase where no low-temperature stable compounds were found. Red squares mark systems for which low-temperature stable compounds are found in computation but no compounds are reported in experiment. Bottom panel: $T_s$ for the binary systems in this work. Colors: from red (lowest $T_s$) to blue (highest $T_s$).

The bottom panel of Fig. 1 ranks systems by their estimated entropic temperature $T_s$. Essentially, the (top panel) map, incorporating the computational data, corresponds to what would be observed at low temperatures, assuming thermodynamic equilibrium, whereas a map with only experimental data reports systems as compound-forming when reaching thermodynamic equilibrium is presumably easier. That is not to say, however, that the predicted phases will necessarily be difficult to synthesize— some of the systems where the $T_s$ value is small have been experimentally observed to be compound-forming (e.g., Cr-Pd, Au-Pd, Ag-Pt, Hg-Rh and Co-Pt). $T_s$ decreases gradually as we move from the centers of the compound-forming clusters towards their edges. Most systems with low $T_s$ are adjacent to the remaining non-compound-forming region. This leads to a qualitative picture of compound stability against disorder which is correlated with the position of a system within the compound forming cluster, and with larger clusters centered at systems with more stable structures.

It is instructive to note that many obscure and large unit cell structures that are reported in the experimental literature turn up in the HT search. For example, compounds of prototypes such as $Mg_{44}Rh_7$, $Ru_{25}Y_{44}$, $Ir_4Sc_{11}$, $Rh_{13}Sc_{57}$ from the experimental literature, nearly always turn up as ground states, or very close to the convex hull, in the HT search as well. This is strong evidence that the first-principles HT approach is robust and has the necessary accuracy to extend the PGM data where experimental results are sparse or difficult to obtain. Also of interest is the appearance of some rare prototypes in systems similar to those in which they were identified experimentally. For example, the prototype $Pd_3Ti_2$, reported only in the Pd-Ti system [36], also emerges as a calculated groundstate in the closely related systems Hf-Pd and Pt-Ti. In Hf-Pt, it appears as marginally stable, at 3meV/atom above the convex hull, in agreement with a very recent experimental study that identified the previously incorrectly charac-ized structure of a $Hf_2Pt_3$ phase [46].

TABLE I: Compounds observed in experiments ("Exper.") or predicted by *ab initio* calculations ("Calc.") in **Osmium** binary alloys (structure prototype in parentheses, multiple entries denote different reported structures, in the experiments, or degenerate structures, in the calculations). "-" denotes no compounds. The superscript "⋆" denotes unobserved prototypes found in calculations [2, 13, 25, 27, 29, 31]. $\Delta H$ are the formation enthalpies from the present study. The energy difference between reported and calculated structures or between the reported structure (unstable in the calculation) and a two-phase tie-line is indicated in brackets "⟨·⟩".

| | Compounds | | $\Delta H$ |
|---|---|---|---|
| | Exper.[35, 36] | Calc. | meV/at. |
| Y | $Os_2Y(C14)$ | $Os_2Y(C14)$ | -304 |
| | $OsY_3(D0_{11})$ | $OsY_3(D0_{11})$ | -239 |
| Sc | $Os_2Sc(C14)$ | $Os_2Sc(C14)$ | -390 |
| | | $OsSc_2(fcc^{[001]}_{AB2})$ | -400 |
| | $Os_4Sc_{11}(Ir_4Sc_{11})$ | $Os_4Sc_{11}(Ir_4Sc_{11})$ | -372 |
| | $Os_7Sc_{44}(Mg_{44}Rh_7)$ | $Os_7Sc_{44}(Mg_{44}Rh_7)$ | -197 |
| Zr | $Os_2Zr(C14)$ | $Os_2Zr(C14)$ | -388 |
| | $OsZr(B2)$ | $OsZr(B2)$ | -524 |
| | $Os_4Zr_{11}(Ir_4Sc_{11})$ | $Os_4Zr_{11}(Ir_4Sc_{11})$ | -29 |
| | $Os_{17}Zr_{54}(Hf_{54}Os_{17})$ | | ⟨8⟩ |
| | | $OsZr_4(D1_a)$ | -220 |
| Hf | $Hf_{54}Os_{17}(Hf_{54}Os_{17})$ | | ⟨20⟩ |
| | $Hf_2Os(NiTi_2)$ | | ⟨44⟩ |
| | $HfOs(B2)$ | $HfOs(B2)$ | -709 |
| | $HfOs_2(C14)$ | | ⟨66⟩ |
| Ti | $OsTi(B2)$ | $OsTi(B2)$ | -714 |
| | | $OsTi_2(C49)$ | -715 |
| | | $OsTi_3(Mo_3Ti^\star)$ | -403 |
| Nb | | $Nb_5Os(HfPd_5^\star)$ | -200 |
| | $Nb_3Os(A15)$ | $Nb_3Os(A15)$ | -275 |
| | $Nb_{0.6}Os_{0.4}(\sigma)$ | $Nb_{20}Os_{10}(\sigma_{BAAABA})$ | -274 |
| | $Nb_{0.4}Os_{0.6}(\chi)$ | $Nb_{12}Os_{17}(Al_{12}Mg_{17})$ | -247 |
| | | $NbOs_3(D0_{24})$ | -115 |
| Ta | | $Os_2Ta(Ga_2Hf)$ | -205 |
| | $Os_{0.5}Ta_{0.5}(\chi)$ | $Os_{12}Ta_{17}(Al_{12}Mg_{17})$ | -313 |
| | $Os_{0.3}Ta_{0.7}(\sigma)$ | $Os_{10}Ta_{20}(\sigma_{ABBAB})$ | -335 |
| | | $OsTa_3(A15)$ | -330 |
| V | | $Os_3V(Re_3Ru^\star)$ | -150 |
| | | $Os_3V_5(Ga_3Pt_5)$ | -350 |
| | | $OsV_2(C11_b)$ | -354 |
| | $OsV_3(A15)$ | $OsV_3(D0_3)$ | -361⟨21⟩ |
| | | $OsV_5(Mo_5Ti^\star)$ | -253 |
| Mo | $Mo_3Os(A15)$ | | ⟨29⟩ |
| | $Mo_{0.65}Os_{0.35}(\sigma)$ | | |
| | | $MoOs_3(D0_{19})$ | -52 |

| | Compounds | | $\Delta H$ |
|---|---|---|---|
| | Exper.[35, 36] | Calc. | meV/at. |
| W | | $Os_3W(D0_{19})$ | -56 |
| | $Os_{0.3}W_{0.7}(\sigma)$ | | |
| Cr | $Cr_3Os(A15)$ | | ⟨18⟩ |
| | | $CrOs_3(D0_{19})$ | -22 |
| | | $CrOs_5(Hf_5Sc^\star)$ | -19 |
| Tc | - | $Os_3Tc(D0_{19})$ | -71 |
| | | $OsTc(B19)$ | -83 |
| | | $OsTc_3(D0_{19})$ | -57 |
| Re | - | $Os_3Re(D0_{19})$ | -78 |
| | | $OsRe(B19)$ | -89 |
| | | $OsRe_2(Sc_2Zr^\star)$ | -68 |
| | | $OsRe_3(Re_3Ru^\star)$ | -56 |
| Mn | - | $MnOs(B19)$ | -42 |
| | | $MnOs_3(D0_{19})$ | -36 |
| Fe | - | - | |
| Os | reference | | |
| Ru | | $Os_3Ru(D0_a)$ | -9 |
| | | $OsRu(B19)$ | -15 |
| | | $OsRu_3(D0_a)$ | -11 |
| | | $OsRu_5(Hf_5Sc^\star)$ | -9 |
| Co | - | - | |
| Ir | - | $Ir_8Os(Pt_8Ti)$ | -8 |
| | | $IrOs_5(Hf_5Sc^\star)$ | -7 |
| Rh | - | $OsRh(RhRu^\star)$ | -8 |
| Ni | - | - | |
| Pt | - | - | |
| Pd | - | - | |
| Au | - | - | |
| Ag | - | - | |
| Cu | - | - | |
| Hg | - | - | |
| Cd | - | - | |
| Zn | - | - | |

In the systems we examined, there are nearly 50 phases reported in the experimental phase diagrams for which the crystal structure of the phase is not known. In one half of these cases, the HT calculations identify stable structures for these unknown phases. For the other half of these unknown structures, our calculations find no stable compounds at the reported concentration, but stable compounds at other concentrations. The reported phases (sans structural information) may, therefore, be due to phases that decompose at low temperatures or may merely represent samples that were kinetically inhibited and unable to settle into their stable phases during the time frame of the experiments.

The prototype database included in this study comprise both experimentally-reported structures as well as hypothetical structures constructed combinatorially from derivative supercells of fcc, bcc, and hcp lattices [37, 47]. Occasionally these derivative superstructures are predicted to be ground states by the first-principles calcu-lations. In this work, we find compounds with 5 of these new structures, for which no prototype is known and no *Strukturbericht* designation have been given. These new prototypes are marked by a † in tables I to VI and their crystallographic parameters are given in Table VII. We also find a few other compounds with unobserved prototypes (marked by a ⋆ in tables I to VI) previously uncovered in related HT studies [2, 13, 25, 27, 29, 31].

TABLE II: Compounds in **Ruthenium** binary alloys. (Unkn.) denotes an unknown structure. All other symbols are as in Table I.

| | Compounds | | $\Delta H$ | | Compounds | | $\Delta H$ |
|---|---|---|---|---|---|---|---|
| | Exper.[35, 36] | Calc. | meV/at. | | Exper.[35, 36] | Calc. | meV/at. |
| Y | Ru₂Y(C14) | Ru₂Y(C14) | -313 | Mo | Mo₀.₆Ru₀.₄(σ) | Mo₁₄Ru₁₆(σ_AABAB) | -116 |
| | Ru₂Y₃(Er₃Ru₂) | | ⟨79⟩ | W | Ru₀.₄W₀.₆(σ) | Ru₃W(D0₁₉) | -65 |
| | Ru₂₅Y₄₄(Ru₂₅Y₄₄) | Ru₂₅Y₄₄(Ru₂₅Y₄₄) | -342 | Cr | Cr₀.₇Ru₀.₃(σ) | - | |
| | Ru₂Y₅(C₂Mn₅) | Ru₂Y₅(C₂Mn₅) | -334 | Tc | | Ru₃Tc(D0₁₉) | -63 |
| | RuY₃(D0₁₁) | RuY₃(D0₁₁) | -307 | | | RuTc(B19) | -73 |
| Sc | Ru₂Sc(C14) | Ru₂Sc(C14) | -389 | | | RuTc₃(D0₁₉) | -47 |
| | RuSc(B2) | RuSc(B2) | -540 | | | RuTc₅(RuTc₅*) | -32 |
| | Ru₃Sc₅(D8₈) | | ⟨42⟩ | Re | - | Re₃Ru(Re₃Ru*) | -53 |
| | RuSc₂(NiTi₂) | RuSc₂(C11_b) | -484⟨84⟩ | | | ReRu(B19) | -86 |
| | Ru₄Sc₁₁(Ir₄Sc₁₁) | Ru₄Sc₁₁(Ir₄Sc₁₁) | -405 | | | ReRu₃(D0₁₉) | -80 |
| | Ru₁₃Sc₅₇(Rh₁₃Sc₅₇) | | ⟨10⟩ | Mn | - | Mn₂₄Ru₅(Re₂₄Ti₅) | -18 |
| | Ru₇Sc₄₄(Mg₄₄Rh₇) | Ru₇Sc₄₄(Mg₄₄Rh₇) | -226 | Fe | - | - | |
| Zr | RuZr(B2) | RuZr(B2) | -644 | Os | - | Os₃Ru(D0_a) | -9 |
| Hf | HfRu(B2) | HfRu(B2) | -819 | | | OsRu(B19) | -15 |
| | HfRu₂(Unkn.) | | | | | OsRu₃(D0_a) | -11 |
| Ti | RuTi(B2) | RuTi(B2) | -763 | | | OsRu₅(Hf₅Sc*) | -9 |
| | | RuTi₂(C49) | -532 | Ru | Reference | | |
| | | RuTi₃(Mo₃Ti*) | -401 | Co | - | - | |
| Nb | | Nb₈Ru(Pt₈Ti) | -117 | Ir | - | Ir₈Ru(Pt₈Ti) | -20 |
| | | Nb₅Ru(Nb₅Ru*) | -172 | | | Ir₃Ru(L1₂) | -34 |
| | | Nb₄Ru(L6₀) | -222 | | | IrRu(B19) | -49 |
| | | Nb₅Ru₃(Ga₃Pt₅) | -249 | | | IrRu₂(Ir₂Tc*) | -54 |
| | NbRu(Unkn.) | | | | | IrRu₃(D0₁₉) | -53 |
| | | Nb₃Ru₅(Ga₃Pt₅) | -240 | | | IrRu₅(Hf₅Sc*) | -37 |
| | NbRu₃(L1₂) | | ⟨8⟩ | Rh | - | Rh₈Ru(Pt₈Ti) | -2 |
| Ta | Ru₅Ta₃(Unkn.) | Ru₅Ta₃(Ga₃Pt₅) | -332 | | | RhRu(RhRu*) | -8 |
| | RuTa(Unkn.) | | | | | RhRu₂(RhRu₂*) | -6 |
| | | Ru₃Ta₅(Ga₃Pt₅) | -313 | | | RhRu₅(RhRu₅*) | -3 |
| | | RuTa₃(fcc_AB3^[001]) | -281 | Ni | - | - | |
| | | RuTa₅(Nb₅Ru*) | -207 | Pt | - | PtRu(CdTi) | -33 |
| V | | Ru₃V(Re₃Ru*) | -145 | Pd | - | - | |
| | | Ru₂V(C37) | -192 | Au | - | - | |
| | RuV(B11) | | ⟨28⟩ | Ag | - | - | |
| | | Ru₃V₅(Ga₃Pt₅) | -313 | Cu | - | - | |
| | | RuV₂(C11_b) | -321 | Hg | - | - | |
| | | RuV₃(Mo₃Ti*) | -296 | Cd | - | - | |
| | | RuV₄(D1_a) | -262 | Zn | | RuZn₃(L1₂) | -150 |
| | | RuV₅(Nb₅Ru*) | -230 | | RuZn₆(RuZn₆) | RuZn₆(RuZn₆) | -132 |
| | | RuV₈(Pt₈Ti) | -154 | | | | |

## IV. STRUCTURE MAPS

Empirical structure maps present available experimental data in ways that highlight similarities in materials behavior in alloy systems. Their arrangement principles usually depend on simple parameters, e.g., atomic number, atomic radius, electronegativity, ionization energy, melting temperature or enthalpy. Several well-known classification methods include Hume-Rothery rules

TABLE III: Compounds in **Iridium** binary alloys. The superscript "§" denotes relaxation of one prototype into another and a "†" denotes new prototypes described in Table VII. The other symbols are as in Table II.

| | Compounds Exper.[35, 36] | Compounds Calc. | $\Delta H$ meV/at. |
|---|---|---|---|
| Y | $Ir_3Y(PuNi_3)$ | | $\langle 21\rangle$ |
| | $Ir_2Y(C15)$ | $Ir_2Y(C15)$ | -803 |
| | $IrY(B2)$ | $IrY(B2)$ | -787 |
| | $Ir_2Y_3(Rh_2Y_3)$ | | $\langle 12\rangle$ |
| | $Ir_3Y_5(Pu_5Rh_3)$ | $Ir_3Y_5(Pu_5Rh_3)$ | -772 |
| | $Ir_2Y_5(C_2Mn_5)$ | $Ir_2Y_5(C_2Mn_5)$ | -640 |
| | $IrY_3(D0_{11})$ | $IrY_3(D0_{11})$ | -564 |
| Sc | | $Ir_7Sc(CuPt_7)$ | -352 |
| | $Ir_3Sc(L1_2)$ | | $\langle 7\rangle$ |
| | $Ir_2Sc(C15)$ | $Ir_2Sc(C14)$ | -783$\langle 35\rangle$ |
| | $IrSc(B2)$ | $IrSc(B2)$ | -1032 |
| | $IrSc_2(NiTi_2)$ | | $\langle 26\rangle$ |
| | $Ir_4Sc_{11}(Ir_4Sc_{11})$ | $Ir_4Sc_{11}(Ir_4Sc_{11})$ | -686 |
| | $Ir_{13}Sc_{57}(Rh_{13}Sc_{57})$ | | $\langle 2\rangle$ |
| | $Ir_7Sc_{44}(Mg_{44}Rh_7)$ | $Ir_7Sc_{44}(Mg_{44}Rh_7)$ | -369 |
| Zr | $Ir_3Zr(L1_2)$ | $Ir_3Zr(L1_2)$ | -709 |
| | $Ir_2Zr(C15)$ | $Ir_2Zr(Ga_2Hf)$ | -766$\langle 87\rangle$ |
| | $IrZr(NiTi)$ | $IrZr(NiTi)$ | -830 |
| | $Ir_3Zr_5(Ir_3Zr_5)$ | $Ir_3Zr_5(Ir_3Zr_5)$ | -732 |
| | $IrZr_2(C16)$ | $IrZr_2(C37)$ | -668$\langle 13\rangle$ |
| | $IrZr_3(SV_3)$ | $IrZr_3(SV_3)$ | -519 |
| Hf | $Hf_2Ir(NiTi_2)$ | $Hf_2Ir(C37)$ | -750$\langle 31\rangle$ |
| | $Hf_5Ir_3(D8_8 / Ir_5Zr_3)$ | $Hf_5Ir_3(Ir_5Zr_3)$ | -814$\langle 14\rangle$ |
| | $HfIr$ (Unkn.) | $HfIr(B27)$ | -949 |
| | | $HfIr_2(Ga_2Hf)$ | -872 |
| | $HfIr_3(L1_2)$ | $HfIr_3(L1_2)$ | -800 |
| Ti | | $Ir_7Ti(CuPt_7)$ | -369 |
| | $Ir_3Ti(L1_2)$ | $Ir_3Ti(L1_2)$ | -716 |
| | | $Ir_2Ti(Ga_2Hf)$ | -779 |
| | | $Ir_5Ti_3(Ga_3Pt_5)$ | -809 |
| | $IrTi$ (Unkn.) | $IrTi(L1_0)$ | -847 |
| | | $IrTi_2(C11_b)$ | -712 |
| | $IrTi_3(A15)$ | $IrTi_3(A15)$ | -566 |
| Nb | $Ir_3Nb(L1_2)$ | $Ir_3Nb(Co_3V)$ | -628$\langle 9\rangle$ |
| | $IrNb(L1_0 / IrTa)$ | $IrNb(L1_0)$ | -542$\langle 2\rangle$ |
| | $Ir_{0.37}Nb_{0.63}(\sigma)$ | $Ir_2Nb_5$ $(\sigma_{ABBAB})$ | -484 |
| | $IrNb_3(A15)$ | $IrNb_3(A15)$ | -433 |
| Ta | $Ir_3Ta(L1_2)$ | $Ir_3Ta(Co_3V)$ | -688$\langle 2\rangle$ |
| | | $Ir_2Ta(Ga_2Hf)$ | -659 |
| | $IrTa(L1_0 / IrTa)$ | $IrTa(L1_0)$ | -594$\langle 3\rangle$ |
| | $Ir_{0.25}Ta_{0.75}(\sigma)$ | $Ir_{10}Ta_{20}(\sigma_{ABBAB})$ | -528 |
| | | $IrTa_3(A15)$ | -479 |
| V | $Ir_3V(L1_2)$ | $Ir_3V(D0_{19})$ | -505$\langle 21\rangle$ |
| | $IrV(IrV / L1_0)$ | $IrV(L1_0)$ | -500§ |
| | $IrV_3(A15)$ | $IrV_3(A15)$ | -497 |
| | | $IrV_8(Pt_8Ti)$ | -225 |
| Mo | $Ir_3Mo(D0_{19})$ | $Ir_3Mo(D0_{19})$ | -332 |
| | | $Ir_2Mo(C37)$ | -337 |
| | $IrMo(B19)$ | $IrMo(B19)$ | -321 |
| | | $IrMo_3(A15)$ | $\langle 75\rangle$ |

| | Compounds Exper.[35, 36] | Compounds Calc. | $\Delta H$ meV/at. |
|---|---|---|---|
| W | | $Ir_8W(Pt_8Ti)$ | -157 |
| | $Ir_3W(D0_{19})$ | $Ir_3W(D0_{19})$ | -350 |
| | | $Ir_2W(C37)$ | -352 |
| | $IrW(B19)$ | $IrW(B19)$ | -300 |
| Cr | $Cr_3Ir(A15)$ | | $\langle 48\rangle$ |
| | $Cr_{0.5}Ir_{0.5}(Mg)$ | $CrIr(B19)$ | -239 |
| | | $CrIr_2(C37)$ | -233 |
| | | $CrIr_3(D0_{19})$ | -228 |
| Tc | - | $Ir_8Tc(Pt_8Ti)$ | -89 |
| | | $Ir_2Tc(Ir_2Tc^\star)$ | -224 |
| | | $IrTc(B19)$ | -287 |
| | | $IrTc_3(D0_{19})$ | -217 |
| Re | - | $Ir_8Re(Pt_8Ti)$ | -94 |
| | | $Ir_2Re(Ir_2Tc^\star)$ | -227 |
| | | $IrRe(B19)$ | -274 |
| | | $IrRe_3(D0_{19})$ | -209 |
| Mn | | $Ir_3Mn(L1_2)$ | -173 |
| | $IrMn(L1_0)$ | $IrMn(B19)$ | -204$\langle 58\rangle$ |
| | | $IrMn_2(C37)$ | -175 |
| | $IrMn_3(L1_2)$ | $IrMn_3(L6_0)$ | -156$\langle 108\rangle$ |
| Fe | | $Fe_3Ir(L6_0)$ | -44 |
| | $Fe_{0.6}Ir_{0.4}(Mg)$ | $FeIr(NbP)$ | -57 |
| | | $FeIr_3(D0_{22})$ | -63 |
| Os | - | $Ir_8Os(Pt_8Ti)$ | -8 |
| | | $IrOs_5(Hf_5Sc^\star)$ | -7 |
| Ru | - | $Ir_8Ru(Pt_8Ti)$ | -20 |
| | | $Ir_3Ru(L1_2)$ | -34 |
| | | $IrRu(B19)$ | -49 |
| | | $IrRu_2(Ir_2Tc^\star)$ | -54 |
| | | $IrRu_3(D0_{19})$ | -53 |
| | | $IrRu_5(Hf_5Sc^\star)$ | -37 |
| Co | - | - | |
| Ir | Reference | | |
| Rh | - | $Ir_3Rh(fcc^{[113]}_{AB3})$ | -15 |
| | | $Ir_2Rh(Pd_2Ti)$ | -20 |
| | | $IrRh(fcc^{[113]}_{A2B2})$ | -21 |
| | | $IrRh_2(Pd_2Ti)$ | -18 |
| Ni | - | $IrNi(NbP)$ | -38 |
| Pt | - | - | |
| Pd | - | - | |
| Au | | | |
| Ag | | | |
| Cu | | | |
| Hg | | | |
| Cd | | | |
| Zn | | $IrZn(IrZn^\dagger)$ | -195 |
| | | $IrZn_2(C49)$ | -238 |
| | | $IrZn_3(NbPd_3)$ | -224 |
| | $Ir_2Zn_{11}(Ir_2Zn_{11})$ | $Ir_2Zn_{11}(Ir_2Zn_{11})$ | -192 |

TABLE IV: Compounds in **Rhodium** binary alloys. All symbols are as in Table III.

| | Compounds | | $\Delta H$ | | Compounds | | $\Delta H$ |
|---|---|---|---|---|---|---|---|
| | Exper.[35, 36] | Calc. | meV/at. | | Exper.[35, 36] | Calc. | meV/at. |
| Y | $Rh_3Y(CeNi_3)$ | $Rh_3Y(CeNi_3)$ | -569 | W | $Rh_{0.8}W_{0.2}(Mg)$ | $Rh_8W(Pt_8Ti)$ | -140 |
| | $Rh_2Y(C15)$ | $Rh_2Y(C15)$ | -742 | | $Rh_3W(D0_{19})$ | $Rh_3W(D0_{19})$ | -274 |
| | $RhY(B2)$ | $RhY(B2)$ | -863 | | | $Rh_2W(C37)$ | -264 |
| | $Rh_2Y_3(Rh_2Y_3)$ | | $\langle 8 \rangle$ | Cr | $Cr_3Rh(A15)$ | | $\langle 103 \rangle$ |
| | $Rh_3Y_5(Unkn.)$ | $Rh_3Y_5(Pu_5Rh_3)$ | -727 | | | $CrRh_2(C37)$ | -117 |
| | $Rh_3Y_7(Fe_3Th_7)$ | $Rh_3Y_7(Fe_3Th_7)$ | -606 | | $CrRh_3(L1_2)$ | $CrRh_3(L1_2)$ | -128 |
| | $RhY_3(D0_{11})$ | $RhY_3(D0_{11})$ | -517 | | | $CrRh_7(CuPt_7)$ | -65 |
| Sc | | $Rh_7Sc(CuPt_7)$ | -348 | Tc | - | $Rh_2Tc(Ir_2Tc^\star)$ | -157 |
| | $Rh_3Sc(L1_2)$ | $Rh_3Sc(L1_2)$ | -620 | | | $RhTc(B19)$ | -175 |
| | $RhSc(B2)$ | $RhSc(B2)$ | -1035 | | | $RhTc_3(D0_{19})$ | -158 |
| | | $Ir_4Sc_{11}(Ir_4Sc_{11})$ | -582 | Re | - | $Re_3Rh(D0_{19})$ | -163 |
| | $Rh_{13}Sc_{57}(Rh_{13}Sc_{57})$ | $Rh_{13}Sc_{57}(Rh_{13}Sc_{57})$ | -424 | | | $ReRh(B19)$ | -184 |
| | | $Ir_7Sc_{44}(Mg_{44}Rh_7)$ | -319 | | | $ReRh_2(Ir_2Tc^\star)$ | -173 |
| Zr | $Rh_3Zr(L1_2)$ | $Rh_3Zr(L1_2)$ | -687 | Mn | $Mn_3Rh(L1_2)$ | | $\langle 153 \rangle$ |
| | $Rh_5Zr_3(Pu_3Pd_5)$ | $Rh_5Zr_3(Pu_3Pd_5)$ | -811 | | $MnRh(B2)$ | $MnRh(B2)$ | -190 |
| | $Rh_4Zr_3(Unkn.)$ | | | | | $MnRh_3(L1_2)$ | -126 |
| | $RhZr(NiTi)$ | $RhZr(B33)$ | -790$\langle 3 \rangle$ | | | $MnRh_7(CuPt_7)$ | -66 |
| | $RhZr_2(NiTi_2 / C16)$ | $RhZr_2(C11_b)$ | -568$\langle 34, 11 \rangle$ | Fe | | $Fe_3Rh(bcc_{AB3}^{[001]})$ | -49 |
| | $RhZr_3(D0_e)$ | $IrZr_3(SV_3)$ | -428§ | | | $Fe_2Rh(Fe_2Rh^\dagger)$ | -57 |
| Hf | $Hf_2Rh(NiTi)$ | $Hf_2Rh(CuZr_2)$ | -633$\langle 13 \rangle$ | | $FeRh(B2)$ | | $\langle 1 \rangle$ |
| | $HfRh(B2)$ | $HfRh(B27)$ | -898$\langle 29 \rangle$ | | | $FeRh_3(D0_{24})$ | -56 |
| | $Hf_3Rh_4(Unkn.)$ | | | Os | - | $OsRh(RhRu^\star)$ | -8 |
| | $Hf_3Rh_5(Ge_3Rh_5)$ | $Hf_3Rh_5(Ge_3Rh_5)$ | -928 | Ru | - | $Rh_8Ru(Pt_8Ti)$ | -2 |
| | $HfRh_3(L1_2)$ | $HfRh_3(L1_2)$ | -762 | | | $RhRu(RhRu^\star)$ | -8 |
| Ti | | $Rh_7Ti(CuPt_7)$ | -330 | | | $RhRu_2(RhRu_2^\star)$ | -6 |
| | $Rh_5Ti(Unkn.)$ | | | | | $RhRu_5(RhRu_5^\star)$ | -3 |
| | $Rh_3Ti(L1_2)$ | $Rh_3Ti(L1_2)$ | -631 | Co | - | - | |
| | $Rh_5Ti_3(Ge_3Rh_5)$ | $Rh_5Ti_3(Ge_3Rh_5)$ | -790 | Ir | - | $Ir_3Rh(fcc_{AB3}^{[113]})$ | -15 |
| | $RhTi (Unkn.)$ | $RhTi(L1_0)$ | -749 | | | $Ir_2Rh(Pd_2Ti)$ | -20 |
| | $RhTi_2(CuZr_2)$ | $RhTi_2(C11_b)$ | -629$\langle 1 \rangle$ | | | $IrRh(fcc_{A2B2}^{[113]})$ | -21 |
| Nb | | $Nb_8Rh(Pt_8Ti)$ | -131 | | | $IrRh_2(Pd_2Ti)$ | -18 |
| | $Nb_3Rh(A15)$ | $Nb_3Rh(A15)$ | -288 | Rh | reference | | |
| | $Nb_{0.7}Rh_{0.3}(\sigma)$ | $Nb_{20}Rh_{10}(\sigma_{BAABA})$ | -342 | Ni | - | - | |
| | $NbRh(L1_0 / IrTa)$ | $NbTh(L1_0)$ | -436$\langle 4 \rangle$ | Pt | - | $PtRh(NbP)$ | -25 |
| | $NbRh_3(L1_2 / Co_3V)$ | $NbRh_3(Co_3V)$ | -548$\langle 6 \rangle$ | | | $PtRh_2(Pd_2Ti)$ | -21 |
| Ta | $Rh_3Ta(L1_2)$ | $Rh_3Ta(L1_2)$ | -611 | | | $PtRh_3(D0_{22})$ | -18 |
| | $Rh_2Ta(C37)$ | $Rh_2Ta(Ga_2Hf)$ | -597$\langle 13 \rangle$ | Pd | - | - | |
| | $RhTa(IrTa)$ | | | Au | - | - | |
| | $Rh_{0.3}Ta_{0.7}(\sigma)$ | $RhTa_3(A15)$ | -333 | Ag | - | - | |
| | | $RhTa_5(RuTc_5^\star)$ | -233 | Cu | - | $Cu_7Rh(CuPt_7)$ | -4 |
| | | $RhTa_8(Pt_8Ti)$ | -159 | Hg | "$Hg_5Rh$" | $Hg_4Rh(Hg_4Pt)$ | -40 |
| V | | $Rh_5V(HfPd_5^\star)$ | -268 | | | | |
| | $Rh_3V(L1_2)$ | $Rh_3V(D0_{19})$ | -393$\langle 11 \rangle$ | | "$Hg_{4.63}Rh$" | | $\langle 28 \rangle$ |
| | $RhV(IrV / L1_0)$ | $RhV(L1_0)$ | -371§ | | $Hg_2Rh(Hg_2Pt)$ | | |
| | $RhV_3(A15)$ | $RhV_3(A15)$ | -332 | Cd | "$Cd_{21}Rh_5$" | $Cd_4Rh(Hg_4Pt)$ | -104 |
| | | $RhV_5(RuTc_5^\star)$ | -246 | | ($\gamma$-brass) | $Cd_2Rh(Hg_2Pt)$ | -166 |
| | | $RhV_8(Pt_8Ti)$ | -170 | Zn | "$Rh_5Zn_{21}$" | $RhZn(B2)$ | -391 |
| Mo | $MoRh(B19)$ | $MoRh(B19)$ | -196 | | ($\gamma$-brass) | $Rh_3Zn_5(Ga_3Pt_5)$ | -395 |
| | | $MoRh_2(C37)$ | -247 | | | $RhZn_2(ZrSi_2)$ | -388 |
| | $MoRh_3(D0_{19})$ | $MoRh_3(D0_{19})$ | -248 | | | $RhZn_3(D0_{23})$ | -351 |
| | | $MoRh_8(Pt_8Ti)$ | -116 | | | | |

TABLE V: Compounds in **Platinum** binary alloys. tet-L1$_2$ denotes a tetragonal distortion of the L1$_2$ structure. All symbols are as in Table III.

| | Compounds | | $\Delta H$ |
| --- | --- | --- | --- |
| | Exper.[35, 36] | Calc. | meV/at. |
| Y | Pt$_5$Y (Unkn.) | Pt$_5$Y(D2$_d$) | -677 |
| | Pt$_3$Y(L1$_2$) | Pt$_3$Y(L1$_2$) | -983 |
| | Pt$_2$Y(C15) | Pt$_2$Y(C15) | -1095 |
| | Pt$_4$Y$_3$ (Unkn.) | | |
| | PtY(B27) | PtY(B33) | -1252(54) |
| | Pt$_4$Y$_5$(Ge$_4$Sm$_5$) | | (1) |
| | Pt$_3$Y$_5$(Mn$_5$Si$_3$) | | (28) |
| | PtY$_2$(Cl$_2$Pb) | PtY$_2$(Cl$_2$Pb) | -936 |
| | Pt$_3$Y$_7$(Fe$_3$Th$_7$) | | (19) |
| | PtY$_3$(D0$_{11}$) | PtY$_3$(D0$_{11}$) | -709 |
| Sc | | Pt$_8$Sc(Pt$_8$Ti) | -482 |
| | Pt$_3$Sc(L1$_2$) | Pt$_3$Sc(L1$_2$) | -1050 |
| | | Pt$_2$Sc(Ga$_2$Hf) | -1143 |
| | PtSc(B2) | PtSc(B2) | -1232 |
| | PtSc$_2$(Cl$_2$Pb) | PtSc$_2$(Cl$_2$Pb) | -982 |
| | Pt$_{13}$Sc$_{57}$(Rh$_{13}$Sc$_{57}$) | Pt$_{13}$Sc$_{57}$(Rh$_{13}$Sc$_{57}$) | -571 |
| Zr | Pt$_8$Zr(Pt$_8$Ti) | Pt$_8$Zr(Pt$_8$Ti) | -496 |
| | Pt$_3$Zr(D0$_{24}$ / L1$_2$) | Pt$_3$Zr(D0$_{24}$) | -1031(12) |
| | Pt$_2$Zr(C11$_b$) | | (62) |
| | Pt$_{11}$Zr$_9$(Pt$_{11}$Zr$_9$) | | (73) |
| | PtZr(TlI) | PtZr(B33) | -1087(1) |
| | Pt$_3$Zr$_5$(D8$_8$) | | (25) |
| | PtZr$_2$(NiTi$_2$) | PtZr$_2$(C16) | -759(51) |
| Hf | Hf$_2$Pt(NiTi$_2$) | Hf$_2$Pt(NiTi$_2$) | -786 |
| | HfPt(B2/B33/TlI) | HfPt(B33/TlI) | -1155(165) |
| | HfPt$_3$(L1$_2$/D0$_{24}$) | HfPt$_3$(D0$_{24}$) | -1100(3) |
| | | HfPt$_8$(Pt$_8$Ti) | -528 |
| Ti | Pt$_8$Ti(Pt$_8$Ti) | Pt$_8$Ti(Pt$_8$Ti) | -433 |
| | | Pt$_5$Ti(HfPd$_5^\star$) | -617 |
| | Pt$_3$Ti(D0$_{24}$/L1$_2$) | Pt$_3$Ti(PuAl$_3$) | -864(3, 5) |
| | | Pt$_2$Ti(C49) | -912 |
| | | Pt$_3$Ti$_2$(Pd$_3$Ti$_2$) | -931 |
| | PtTi(B19) | PtTi(NiTi) | -933(5) |
| | PtTi$_3$(A15) | PtTi$_3$(A15) | -648 |
| Nb | Nb$_3$Pt(A15) | Nb$_3$Pt(A15) | -415 |
| | Nb$_{0.6}$Pt$_{0.4}$($\sigma$) | | |
| | NbPt(B19) | NbPt(L1$_0$) | -660(13) |
| | NbPt$_2$(MoPt$_2$) | NbPt$_2$(MoPt$_2$) | -721 |
| | NbPt$_3$(L6$_0$/NbPt$_3$) | NbPt$_3$(NbPt$_3$/D0$_a$) | -678(154) |
| | | NbPt$_8$(Pt$_8$Ti) | -378 |
| Ta | | Pt$_8$Ta(Pt$_8$Ti) | -416 |
| | Pt$_3$Ta(D0$_{22}$/L6$_0$/NbPt$_3$) | Pt$_3$Ta(NbPt$_3$) | -723(11, 183) |
| | Pt$_2$Ta(Au$_2$V) | Pt$_2$Ta(Au$_2$V) | -757 |
| | | PtTa(L1$_0$) | -643 |
| | Pt$_{0.25}$Ta$_{0.75}$($\sigma$) | Pt$_5$Ta$_2$($\sigma_{BBBAB}$) | -434 |
| | Pt$_{0.6}$Ta$_{3.74}$(A15) | PtTa$_3$(A15) | -416 |
| | | PtTa$_8$(Pt$_8$Ti) | -197 |
| V | | Pt$_8$V(Pt$_8$Ti) | -275 |
| | Pt$_3$V(D0$_{22}$) | Pt$_3$V(D0$_a$) | -464(4) |
| | Pt$_2$V(MoPt$_2$) | Pt$_2$V(MoPt$_2$) | -555 |
| | PtV(B19) | PtV(L1$_0$) | -563(2) |
| | PtV$_3$(A15) | PtV$_3$(A15) | -436 |
| | | PtV$_8$(Pt$_8$Ti) | -206 |
| Mo | Mo$_3$Pt(D0$_{19}$) | | (38) |
| | MoPt(B19) | MoPt(B19) | -321 |
| | MoPt$_2$(MoPt$_2$) | MoPt$_2$(MoPt$_2$) | -366 |
| | MoPt$_3$ (Unkn.) | | |
| | | MoPt$_4$(D1$_a$) | -265 |
| | | MoPt$_8$(Pt$_8$Ti) | -180 |
| W | | Pt$_8$W(Pt$_8$Ti) | -202 |
| | | Pt$_4$W(D1$_a$) | -270 |
| | Pt$_2$W(MoPt$_2$) | Pt$_2$W(MoPt$_2$) | -343 |
| | PtW (Unkn.) | | |
| Cr | Cr$_3$Pt(A15 / L1$_2$) | | (70) |
| | CrPt(L1$_0$) | CrPt(B19) | -191(31) |
| | CrPt$_3$(L1$_2$) | CrPt$_3$(L1$_2$) | -261 |
| | | CrPt$_8$(Pt$_8$Ti) | -136 |

| | Compounds | | $\Delta H$ |
| --- | --- | --- | --- |
| | Exper.[35, 36] | Calc. | meV/at. |
| Tc | - | Pt$_3$Tc(bcc$^{[001]}_{AB3}$) | -158 |
| | | Pt$_2$Tc(Ir$_2$Tc$^\star$) | -184 |
| | | PtTc$_3$(bcc$^{[001]}_{AB3}$) | -267 |
| Re | Pt$_8$Re(Unkn.) | Pt$_3$Re(bcc$^{[001]}_{AB3}$) | -128 |
| | | PtRe$_3$(bcc19) | -231 |
| Mn | Mn$_3$Pt(L1$_2$) | Mn$_3$Pt(D0$_{19}$) | -174(144) |
| | MnPt(L1$_0$) | | (293) |
| | | Mn$_3$Pt$_5$(Ga$_3$Pt$_5$) | -363 |
| | | MnPt$_2$(Ga$_2$Hf) | -365 |
| | MnPt$_3$(L1$_2$) | MnPt$_3$(L1$_2$) | -363 |
| | | MnPt$_8$(Pt$_8$Ti) | -172 |
| Fe | Fe$_3$Pt(L1$_2$) | | (39) |
| | FePt(L1$_0$) | FePt(L1$_0$) | -244 |
| | | FePt$_2$(Ga$_2$Hf) | -220 |
| | FePt$_3$(L1$_2$) | FePt$_3$(tet-L1$_2$ $c/a=.992$) | -203 |
| | | FePt$_5$(HfPd$_5^\star$) | -162 |
| Os | - | - | |
| Ru | - | PtRu(CdTi) | -33 |
| Co | Co$_3$Pt(D0$_{19}$) | Co$_3$Pt(D0$_{19}$) | -97 |
| | CoPt(L1$_0$) | | (12) |
| | | CoPt$_2$(CuZr$_2$) | -106 |
| | CoPt$_3$(L1$_2$) | CoPt$_3$(L1$_2$) | (16) |
| | | CoPt$_5$(HfPd$_5^\star$) | -55 |
| Ir | - | - | |
| Rh | | PtRh(NbP) | -25 |
| | | PtRh$_2$(Pd$_2$Ti) | -21 |
| | | PtRh$_3$(D0$_{22}$) | -18 |
| Ni | Ni$_3$Pt (Unkn.) | Ni$_3$Pt(D0$_{22}$) | -76 |
| | NiPt(L1$_0$) | NiPt(L1$_0$) | -99 |
| | | NiPt$_2$(CuZr$_2$) | -75 |
| | | NiPt$_3$(D0$_{23}$) | -61 |
| Pt | reference | | |
| Pd | - | Pd$_7$Pt(CuPt$_7$) | -14 |
| | | Pd$_3$Pt(CdPt$_5^\star$) | -25 |
| | | PdPt(L1$_1$) | -36 |
| | | PdPt$_3$(L1$_2$) | -26 |
| | | PdPt$_7$(CuPt$_7$) | -15 |
| Au | - | - | |
| Ag | | Ag$_7$Pt(CuPt$_7$) | -13 |
| | Ag$_3$Pt(L1$_2$) | | (34) |
| | | Ag$_3$Pt$_2$(Ag$_3$Pt$_2^\ddagger$) | -38 |
| | AgPt(Unkn.) | AgPt(L1$_1$) | -39 |
| | AgPt$_3$ (Unkn.) | | |
| Cu | Cu$_7$Pt(CuPt$_7$) | Cu$_7$Pt(CuPt$_7$) | -87 |
| | Cu$_3$Pt(L1$_2$) | Cu$_3$Pt(L1$_2$) | -143 |
| | CuPt(L1$_1$) | CuPt(L1$_1$) | -166 |
| | Cu$_3$Pt$_5$ (Unkn.) | | |
| | CuPt$_3$(Unkn.) | CuPt$_3$(CdPt$_3^\star$) | -121 |
| | CuPt$_7$(CuPt$_7$) | CuPt$_7$(CuPt$_7$) | -77 |
| Hg | Hg$_4$Pt(Hg$_4$Pt) | Hg$_4$Pt(Hg$_4$Pt) | -104 |
| | Hg$_2$Pt(Hg$_2$Pt) | | (25) |
| | HgPt$_3$ (Unkn.) | | |
| Cd | Cd$_5$Pt(Cd$_5$Pt-partial occupancy) | | |
| | | Cd$_4$Pt(Hg$_4$Pt) | -228 |
| | Cd$_3$Pt (Unkn.) | Cd$_3$Pt(D0$_{11}$) | -260 |
| | Cd$_7$Pt$_3$ (Unkn.) | | |
| | Cd$_2$Pt(Unkn.) | Cd$_2$Pt(Hg$_2$Pt) | -316 |
| | CdPt(L1$_0$) | CdPt(L1$_0$) | -322 |
| | CdPt$_3$(L1$_2$) | CdPt$_3$(CdPt$_3^\star$) | -190(11) |
| | | CdPt$_7$(CuPt$_7$) | -114 |
| Zn | | Pt$_7$Hg(CuPt$_7$) | -189 |
| | Pt$_3$Zn(L1$_2$) | Pt$_3$Zn(CdPt$_5^\star$) | -331(6) |
| | PtZn(L1$_0$) | PtZn(L1$_0$) | -570 |
| | PtZn$_2$(Unkn.) | PtZn$_2$(C49) | -463 |
| | | PtZn$_3$(D0$_{22}$) | -397 |
| | Pt$_2$Zn$_{11}$(Ir$_2$Zn$_{11}$) | | -272 |
| | PtZn$_8$ (Unkn.) | | |

TABLE VI: Compounds in **Palladium** binary alloys. tet-fcc denotes a tetragonal distortion of stacked fcc superstructures. All symbols are as in Table III.

| | Compounds Exper.[35, 36] | Calc. | $\Delta H$ meV/at. |
|---|---|---|---|
| Y | Pd$_7$Y(CuPt$_7$) | Pd$_7$Y(CuPt$_7$) | -442 |
| | Pd$_3$Y(L1$_2$) | Pd$_3$Y(L1$_2$) | -863 |
| | Pd$_2$Y (Unkn.) | | |
| | Pd$_3$Y$_2$ (Unkn.) | | |
| | Pd$_4$Y$_3$(Pd$_4$Pu$_3$) | Pd$_4$Y$_3$(Pd$_4$Pu$_3$) | -923 |
| | PdY (Unkn.) | PdY(B33) | -913 |
| | Pd$_2$Y$_3$(Ni$_2$Er$_3$) | | (8) |
| | | PdY$_2$(C37) | -622 |
| | PdY$_3$(D0$_{11}$) | PdY$_3$(D0$_{11}$) | -475 |
| Sc | | Pd$_8$Sc(Pt$_8$Ti) | -411 |
| | Pd$_3$Sc(L1$_2$) | Pd$_3$Sc(L1$_2$) | -855 |
| | Pd$_2$Sc (Unkn.) | Pd$_2$Sc(C37) | -898 |
| | PdSc(B2) | PdSc(B2) | -906 |
| | PdSc$_2$(NiTi$_2$) | PdSc$_2$(NiTi$_2$) | -660 |
| | PdSc$_4$ (Unkn.) | | |
| Zr | | Pd$_8$Zr(Pt$_8$Ti) | -424 |
| | | Pd$_5$Zr(HfPd$_5^\star$) | -591 |
| | Pd$_3$Zr(D0$_{24}$) | Pd$_3$Zr(D0$_{24}$) | -816 |
| | Pd$_2$Zr(C11$_b$) | | (1) |
| | Pd$_4$Zr$_3$(Pd$_4$Pu$_3$) | | (2) |
| | PdZr (Unkn.) | PdZr(B33) | -645 |
| | Pd$_3$Zr$_5$(D8$_a$) | | (90) |
| | PdZr$_2$(NiTi$_2$ / CuZr$_2$) | PdZr$_2$(C11$_b$/CuZr$_2$) | -487(83) |
| Hf | Hf$_2$Pd(C11$_b$ / CuZr$_2$ ) | Hf$_2$Pd(C11$_b$/CuZr$_2$) | -527 |
| | HfPd (Unkn.) | HfPd(B33) | -685 |
| | | Hf$_2$Pd$_3$(Pd$_3$Ti$_2$) | -778 |
| | Hf$_3$Pd$_4$ (Unkn.) | | |
| | | Hf$_3$Pd$_5$(Pd$_5$Ti$_3$) | -800 |
| | HfPd$_2$(C11$_b$) | | (9) |
| | HfPd$_3$(D0$_{24}$/L1$_2$) | HfPd$_3$(D0$_{24}$) | -879(11) |
| | | HfPd$_5$(HfPd$_5^\star$) | -635 |
| | | HfPd$_8$(Pt$_8$Ti) | -430 |
| Ti | | Pd$_5$Ti(HfPd$_5^\star$) | -481 |
| | Pd$_{3.2}$Ti$_{0.8}$(L1$_2$) | | (7) |
| | Pd$_3$Ti(D0$_{24}$) | Pd$_3$Ti(D0$_{24}$) | -646 |
| | Pd$_2$Ti(Pd$_2$Ti) | Pd$_2$Ti(Pd$_2$Ti) | -632 |
| | Pd$_5$Ti$_3$(Pd$_5$Ti$_3$) | Pd$_5$Ti$_3$(Pd$_5$Ti$_3$) | -615 |
| | Pd$_3$Ti$_2$(Pd$_3$Ti$_2$) | Pd$_3$Ti$_2$(Pd$_3$Ti$_2$) | -602 |
| | PdTi(B19) | | (5) |
| | PdTi$_2$(CuZr$_2$) | PdTi$_2$(C11$_b$/CuZr$_2$) | -451 |
| | Pd$_{0.8}$Ti$_{3.2}$(A15) | PdTi$_3$(A15) | -342 |
| Nb | | Nb$_3$Pd(Nb$_3$Pd$^\dagger$) | -167 |
| | | Nb$_2$Pd(CuZr$_2$) | -220 |
| | Nb$_{0.6}$Pd$_{0.4}$($\sigma$) | | |
| | NbPd$_2$(MoPt$_2$) | NbPd$_2$(MoPt$_2$) | -432 |
| | NbPd$_3$(NbPd$_3$ / D0$_{22}$) | NbPd$_3$(NbPd$_3$) | -435(2) |
| | | NbPd$_5$(HfPd$_5^\star$) | -356 |
| | | NbPd$_8$(Pt$_8$Ti) | -279 |
| Ta | | Pd$_8$Ta(Pt$_8$Ti) | -325 |
| | | Pd$_5$Ta(HfPd$_5^\star$) | -401 |
| | Pd$_3$Ta(D0$_{22}$) | Pd$_3$Ta(D0$_{22}$) | -480 |
| | Pd$_2$Ta(MoPt$_2$) | Pd$_2$Ta(MoPt$_2$) | -458 |
| | PdTa(B11) | PdTa(B11) | -362 |
| | Pd$_{0.25}$Ta$_{0.75}$($\sigma$) | | |
| | | PdTa$_8$(Pt$_8$Ti) | -98 |
| V | | Pd$_8$V(Pt$_8$Ti) | -177 |
| | Pd$_3$V(D0$_{22}$) | Pd$_3$V(NbPd$_3$) | -253(6) |
| | Pd$_2$V(MoPt$_2$) | Pd$_2$V(MoPt$_2$) | -274 |
| | PdV (Unkn.) | | |
| | PdV$_3$(A15) | | (9) |
| | | Pd$_5$V(Mo$_5$Ti$^\star$) | -115 |
| | | PdV$_8$(Pt$_8$Ti) | -94 |
| Mo | MoPd$_2$(MoPt$_2$) | MoPd$_2$(MoPt$_2$) | -99 |
| | | MoPd$_4$(D1$_a$) | -92 |
| | | MoPd$_8$(Pt$_8$Ti) | -86 |
| W | - | Pd$_8$W(Pt$_8$Ti) | -122 |
| Cr | Cr$_{0.49}$Pd$_{0.51}$(In) | | |
| | Cr$_{1.33}$Pd$_{2.67}$(L1$_2$) | CrPd$_3$(L1$_2$) | -81 |
| | | CrPd$_5$(HfPd$_5^\star$) | -76 |
| Tc | - | PdTc(RhRu$^\star$) | -63 |
| | | PdTc$_3$(D0$_{19}$) | -73 |

| | Compounds Exper.[35, 36] | Calc. | $\Delta H$ meV/at. |
|---|---|---|---|
| Re | - | PdRe$_3$(D0$_{19}$) | -56 |
| Mn | | MnPd(L1$_0$) | (5) |
| | Mn$_3$Pd$_5$(Ga$_3$Pt$_5$) | Mn$_3$Pd$_5$(Ga$_3$Pt$_5$) | -250 |
| | | MnPd$_2$(C37) | -252 |
| | MnPd$_3$(D0$_{23}$) | MnPd$_3$(L1$_2$) | -234(10) |
| | | MnPd$_5$(HfPd$_5^\star$) | -175 |
| | | MnPd$_8$(Pt$_8$Ti) | -125 |
| Fe | Fe$_{0.96}$Pd$_{1.04}$(L1$_0$) | | (22) |
| | | FePd$_2$(CuZr$_2$) | -116 |
| | FePd$_3$(L1$_2$) | FePd$_3$(D0$_{23}$) | -112(2) |
| | | FePd$_5$(HfPd$_5^\star$) | -100 |
| | | FePd$_8$ (Pt$_8$Ti) | -81 |
| Os | - | | |
| Ru | - | | |
| Co | - | CoPd$_3$(tet-fcc$_{AB3}^{[001]}$ $c/a=2.8$) | -10 |
| Ir | - | | |
| Rh | - | | |
| Ni | - | NiPd$_3$(tet-fcc$_{AB3}^{[001]}$ $c/a=2.7$) | -6 |
| Pt | - | Pd$_7$Pt(CuPt$_7$) | -14 |
| | | Pd$_3$Pt(CdPt$_3^\star$) | -25 |
| | | PdPt(L1$_1$) | -36 |
| | | PdPt$_3$(L1$_2$) | -26 |
| | | PdPt$_7$(CuPt$_7$) | -15 |
| Pd | reference | | |
| Au | | Au$_5$Pd(HfPd$_5^\star$) | -55 |
| | Au$_2$Pd(Unkn.) | Au$_2$Pd(D0$_{23}$) | -82 |
| | | Au$_2$Pd(C49) | -88 |
| | AuPd (Unkn.) | AuPd(NbP) | -94 |
| | AuPd$_3$(Unkn.) | AuPd$_3$(L1$_2$) | -56 |
| Ag | - | Ag$_7$Pd(CuPt$_7$) | -33 |
| | | Ag$_5$Pd(HfPd$_5^\star$) | -41 |
| | | Ag$_3$Pd(D0$_{23}$) | -58 |
| | | Ag$_2$Pd(C37) | -63 |
| | | Ag$_2$Pd$_3$(Ag$_2$Pd$_3^\star$) | -63 |
| | | AgPd(L1$_1$) | -59 |
| | | AgPd$_3$(CdPt$_3^\star$) | -31 |
| Cu | Cu$_7$Pd(CuPt$_7$) | | (18) |
| | | Cu$_5$Pd(HfPd$_5^\star$) | -72 |
| | Cu$_3$Pd(L1$_2$ / SrPb$_3$) | Cu$_3$Pd(D0$_{23}$) | -107(2, 5) |
| | | Cu$_2$Pd(Ga$_2$Hf) | -117 |
| | PdCu (Unkn.) | CuPd(B2) | -125 |
| | | CuPd$_3$(L1$_2$) | -71 |
| | CuPd$_7$(CuPt$_7$) | CuPd$_7$(CuPt$_7$) | -37 |
| Hg | Hg$_4$Pd(Unkn.) | Hg$_4$Pd(Hg$_4$Pt) | -101 |
| | Hg$_5$Pd$_2$(Hg$_5$Mn$_2$) | | (65) |
| | | Hg$_2$Pd(Hg$_2$Pt) | -150 |
| | HgPd(L1$_0$) | HgPd(L1$_0$) | -174 |
| | | Hg$_3$Pd$_5$(Ga$_3$Pt$_5$) | -166 |
| | | Hg$_2$Pd$_3$ | -160 |
| | HgPd$_3$ (Unkn.) | HgPd$_3$(D0$_{22}$) | -139 |
| | | HgPd$_4$(D1$_a$) | -112 |
| Cd | Cd$_{11}$Pd$_2$(Ir$_2$Zn$_{11}$) | Cd$_{11}$Pd$_2$(Ir$_2$Zn$_{11}$) | -171 |
| | Cd$_4$Pd (Unkn.) | | |
| | Cd$_3$Pd (Unkn.) | | |
| | | Cd$_2$Pd(Hg$_2$Pt) | -307 |
| | CdPd(CuTi) | CdPd(L1$_0$) | -418(164) |
| | | CdPd$_2$(C37) | -334 |
| | | CdPd$_3$(D0$_{22}$) | -272 |
| | | CdPd$_4$(D1$_a$) | -225 |
| | | CdPd$_5$(HfPd$_5^\star$) | -188 |
| | | CdPd$_7$(CuPt$_7$) | -142 |
| Zn | | Pd$_8$Zn(Pt$_8$Ti) | -165 |
| | Pd$_2$Zn(C37) | Pd$_2$Zn(C37) | -462 |
| | PdZn(CuTi) | PdZn(L1$_0$) | -570(187) |
| | Pd$_3$Zn$_5$ (Unkn.) | | |
| | | PdZn$_3$(D0$_{22}$) | -359 |
| | Pd$_2$Zn$_{11}$(Ir$_2$Zn$_{11}$) | Pd$_2$Zn$_{11}$(Ir$_2$Zn$_{11}$) | -243 |

minimal

TABLE VII: Geometry of new prototypes marked by † in tables III to VI.

| Formula | IrZn | Nb$_3$Pd | Fe$_2$Rh | Ag$_3$Pt$_2$ | Ag$_2$Pd$_3$ |
|---|---|---|---|---|---|
| Lattice | Monoclinic | Orthorhombic | Orthorhombic | Rhombohedral | Monoclinic |
| Space Group | $C2/m$ No. 12 | $Cmmm$ No. 65 | $Cmmm$ No. 65 | $R\bar{3}m$ No. 166 | $C2/m$ No. 12 |
| Pearson symbol | mS8 | oS8 | oS12 | hR5 | mS10 |
| Bravais lattice type | MCLC | ORCC | ORCC | RHL | MCLC |
| Lattice variation [44] | MCLC$_1$ | ORCC | ORCC | RHL$_1$ | MCLC$_3$ |
| Conv. Cell: $a, b, c$ (Å) | 1.94, 3.83, 1.12 | 1.26, 1.78, 3.56 | 1.78, 5.35, 1.26 | 1.12, 1.12, 13.75 | 3.55, 1.59, 1.94 |
| $\alpha, \beta, \gamma$ (deg) | 72.98, 90, 90 | 90, 90, 90 | 90, 90, 90 | 90, 90, 120 | 65.9, 90, 90 |
| Wyckoff positions [45] | Ir $\frac{1}{6}, \frac{1}{2}, -0.292$ (4i)<br>Zn $\frac{1}{6}, \frac{1}{2}, -0.208$ (4i) | Nb1 0, 0, $\frac{1}{4}$ (4k)<br>Nb2 $\frac{1}{2}$, 0, $\frac{1}{2}$ (2c)<br>Pd $\frac{1}{2}$, 0, 0 (2b) | Fe1 $\frac{1}{8}$, 0, 0 (4g)<br>Fe2 0, 0, $\frac{1}{2}$ (2d)<br>Fe3 $\frac{1}{2}$, 0, 0 (2b)<br>Rh $\frac{3}{8}$, 0, $\frac{1}{2}$ (4h) | Ag1 0, 0, $\frac{1}{5}$ (2c)<br>Ag2 0, 0, 0 (1a)<br>Pt 0, 0, $\frac{2}{5}$ (2c) | Ag $\frac{9}{10}, \frac{1}{2}, \frac{1}{10}$ w (4i)<br>Pd1 0, 0, $\frac{1}{2}$ (2c)<br>Pd2 $\frac{1}{10}, \frac{1}{2}, \frac{7}{10}$ (4i) |
| AFLOW label [33] | 123 | 72 | b83 | f38 | f55 |

[48], Miedema formation enthalpy [49], Zunger pseudopotential radii maps [50], and Pettifor maps [42, 43]. These empirical rules and structure maps have helped direct a few successful searches for previously unobserved compounds [51]. However, they offer a limited response to the challenge of identifying new compounds because they rely on the existence of consistent and reliable experimental input for systems spanning most of the relevant parameter space. In many cases, reliable information is missing in a large portion of this space, e.g. less than 50% of the binary systems have been satisfactorily characterized [52]. This leaves considerable gaps in the empirical structure maps and reduces their predictive usefulness. The advance of HT computational methods makes it possible to fill these gaps in the experimental data with complementary *ab initio* data by efficiently covering extensive lists of candidate structure types [28]. This development was envisioned by Pettifor a decade ago [51], and here we present its realization for PGM alloys.

Fig. 2 shows a Pettifor structure map, enhanced by our HT computational results, for structures of 1:1 stoichiometry. The elements along the map axes are ordered according to Pettifor's chemical scale ($\chi$ parameter) [43]. Circles indicate agreement between computation and experiment, regarding the existence of 1:1 compounds, or lack thereof. If the circle contains a label (Strukturbericht or prototype) this denotes the structure that is stable in the given system at this stoichiometry. Rectangles denote disagreement between experiments and computation about the 1:1 compounds, in systems reported as compound forming (blue rectangles) or as non-compound forming (red and gray rectangles). In the lower left part of the map, there is a region of non-compound forming systems, whereas the upper part of the map is mostly composed of compound-forming systems. In the upper part of the map, experiment and computation agree, preserving a large cluster of B2 structures, or differ slightly on the structure reported to have the lowest formation enthalpy at 1:1 (blue rectangles). For example, the 1:1 phases of Hf-Pd and Pd-Zr are unknown according to the phase diagram literature, but we find the stable phases with B33 structure, right next to Hf-Pt in the diagram, which is reported as a B33 structure. Similarly, stable L1$_0$ structures are identified in the Ir-Ti and Rh-Ti systems, adjacent to a reported cluster of this structure. Two additional L1$_0$ structures are identified in the Cd-Pd and Pd-Zn systems, instead of the reported CuTi structures, extending a small known cluster of this structure at the bottom right corner of the map. These are examples of the capability of HT *ab initio* results to complement the empirical Pettifor maps, and extend their regions of predictive input, in a way consistent with the experimental data.

In the middle of the map, in a rough transition zone between compound-forming and non-compound-forming regions, computation finds quite a few cases where stable compounds are predicted in systems where none have been reported experimentally (pink rectangles). Most prominent here is a large cluster of B19 compounds. Nine systems marked by light gray rectangles are reported in experiments as having no compounds, but our calculations find stable compounds at stoichiometries other than 1:1.

At the stoichiometries of 1:2 and 2:1, Fig. 3 shows significant additions of the calculations to the experimental data on compound-formation. Again, the systems where computation finds stable compounds in experimentally non-compound-forming systems are found at the border between the compound-forming region (dark gray circles and white labeled circles) and the non-compound-forming region (light gray circles), or fill isolated gaps within the compound-forming regions. The calculations augment islands of structurally-similar regions, yielding a more consistent structure map. For example, calculation finds the CuZr$_2$ structure for Nb-Pd, extending the island of this structure already present in the experimental results (left panel, upper right). The calculations significantly extend the Hg$_2$Pt island in the lower right of the B$_2$A panel, from a single experimental entry to 6 systems (in Hg-Pt itself, the calculation finds this structure slightly unstable at $T = 0K$, 25meV/atom above the stability tie-line). A cluster of $\sigma$ phases in the left panel shows that this reported disordered phase has underlying ordered realizations at low temperatures. Three completely new islands, for the C37, Ga$_2$Hf and IrTc$_2$ structures, appear near the upper center of the A$_2$B panel. Another new cluster, of the Pd$_2$Ti structure, appears at the lower center of both panels. In general, the clusters of blue rectangles, show that the calculations augment the experimental results in a consistent manner.

The structure map for 1:3 phases is shown in Fig. 4. Similarly to the 1:1 and 1:2 maps, the calculation extends structural islands of the experimental data, most new phases in non-compound-forming systems occur in systems at the boundary between compound-forming and non-compound-forming regions, and there is significant agreement between the experimentally reported phases (or lack thereof) and calculated phases. In the upper part of the right panel, the $L1_2$ and $D0_{24}$ clusters are preserved with slight modifications at their boundaries (at Pt-Ti, the $PuAl_3$ structure is only 3 meV/atom lower than the experimental structure $D0_{24}$, a difference too small to be significant). The $D0_{19}$ cluster is significantly expanded. In the left panel, the calculations introduce a new $D0_{19}$ island near the center of the diagram. New small regions of the $D0_{22}$ structures emerge at the right bottom of both panels. Adjacent $D0_{23}$ and $CdPt_3^*$ islands appear in the left and right panels, respectively. The experimental $D0_e$ structure for $RhZr_3$ may actually be $SV_3$, since in the calculation the $D0_e$ structure relaxed into the $SV_3$ structure, creating a small $SV_3$ island at the top of the left panel.

The structure maps of figs. 2 to 4 give a bird's eye view of the exhaustive HT search for new structures. Consistently with the empirical maps, they show significant separation of different structures into regions where the constituent elements have a similar Pettifor $\chi$ number. The HT data significantly enhances the empirical maps, extends the regions of some structures, fills in apparent gaps and indicates previously unsuspected structure clusters. Moreover, the HT data contains more detail than is apparent in the structure maps. Even when calculation and experiment agree that a system is compound-forming (green [dark gray] circles in Fig. 1), the calculations often find additional stable compounds, beyond those known in experiment. When the reported structures are found to be unstable in the calculation, they are usually just slightly less stable than the calculated groundstate, or just slightly above the convex hull in a two phase region. Such cases and numerous additional predictions of marginally stable structures harbor further opportunities for materials engineering and applications.

## V. CONCLUSIONS

In this study, the low temperature phase diagrams of all binary PGM-transition metal systems are constructed by HT *ab initio* calculations. The picture of PGM alloys emerging from this study is much more complete than that depicted by current experimental data, with dozens of stable structures that have not been previously reported. We predict ordering in 37 systems reported to be phase-separating and in five systems where only disordered phases are reported. In addition, in the known ordering systems, we find many cases in which more phases are predicted to be stable than reported in the experimental phase diagrams. These *ab initio* results complement the ordering tendencies implied by the empirical Pettifor maps. Augmenting the experimental data compiled in the phase-diagram databases [35, 36] with high-throughput first-principles data [33, 34], we construct Pettifor-type structure maps that point to new opportunities for alloys research. These maps demonstrate that the integration of the empirical and computational data produces enhanced maps that should provide a more comprehensive foundation for rational materials design. The theoretical predictions presented here will hopefully serve as a motivation for their experimental validation and be a guide for future studies of these important systems.

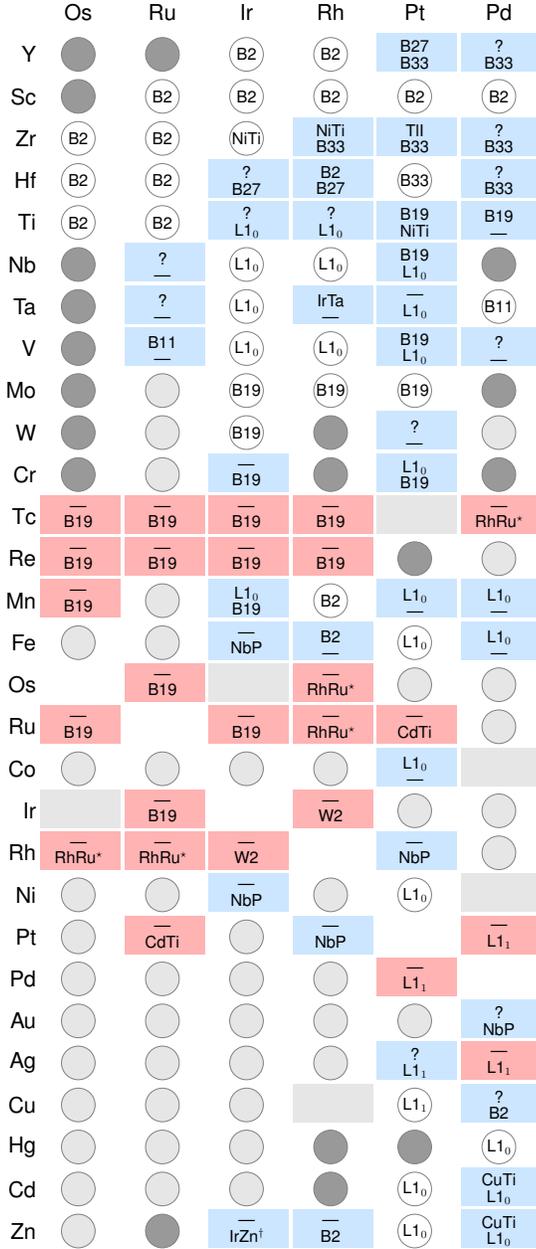

FIG. 2: A Pettifor-type structure map for 1:1 stoichiometry compounds in PGM binary systems. Circles indicate agreement between experiment and computation: white circles with Strukturbericht or prototype labels denote 1:1 compounds, dark circles indicate a compound-forming system with no compounds at 1:1, light circles denote non-compound forming systems. Blue rectangles denote compound-forming systems where the reported and computed stable structures differ at 1:1 stoichiometry. The top label in the rectangle is the reported structure, the bottom label is the structure we find to be stable in this work. A dash "—" indicates the absence of a stable structure. Unidentified suspected structures are denoted by a question mark "?". Pink rectangles indicate systems reported as non-compound forming, with a dash at the top of the rectangle, but we find a stable 1:1 phase, identified at the bottom of the rectangle. Light gray rectangles indicate systems reported as non-compound forming where a structure is predicted at a stoichiometry different from 1:1. A dark gray rectangle indicates a system reported with a disordered compound where no stable structures are found in the calculation.

FIG. 3: A Pettifor-type structure map for 1:2 stoichiometry compounds in PGM binary systems. The symbols are as in Fig. 2, with the map stoichiometry changed respectively from 1:1 to 1:2 or 2:1.

The maps in Figs. 2-4 include a large number of light blue rectangles, pointing to experiment-theory mismatches on structures at simple compositions in binary systems known to be compound forming. This may raise reservations that the level of theory employed, DFT-PBE, may not be as good as commonly accepted for transition metal alloys. A more careful look, however, shows that many of these mismatches, e.g. HfIr, PdZr, Cd$_2$Pt, CuPt$_3$ and Au$_3$Pd, involve cases where a compound of unknown struc-ture has been reported by experiments. The calculation thus reveals the stable structure and closes the gap in the experimental data. In most other cases, e.g. RhZr, PtV, Ir$_3$V, Rh$_2$Ta, Cu$_3$Pt, the energy difference between the reported structure and the calculated structure or two-phase tie-line is rather small and is congruent with the adjacent structure clusters in the maps. Similar improved consistency with reported structure clusters also appears in cases where the discrepancies are considerable, e.g. CdPd and PdZn. In addition, as discussed in Sec. III, the calcula-tions reproduce many complex large unit cell structures that are reported in the experimental literature. Moreover, it is important to remember that experiments are performed at room temperature or higher, while our calculations are carried out at zero temperature. Many phase discrepancies may therefore be due to vibrational promotion [53], or the tendency of structures to gain symmetries by loosing their internal Peierls instabilities or Jahn-Teller distortions. Therefore, the disagreements emerging in our calculations may not be a sign of deficiencies in the theo-retical treatment, but a demonstration of its usefulness is bridging gaps in the experimental data and extending it towards unknown phase transitions at lower temperatures. The ultimate test of this issue rests with experimental validation of at least some of our predictions, which would hopefully be motivated by this work.

| $B_3$\A | Os | Ru | Ir | Rh | Pt | Pd | Os | Ru | Ir | Rh | Pt | Pd | $A_3$/B |
|---|---|---|---|---|---|---|---|---|---|---|---|---|---|
| Y | $D0_{11}$ | $D0_{11}$ | $D0_{11}$ | $D0_{11}$ | $D0_{11}$ | $D0_{11}$ | | | $PuNi_3$ | $Mo_3Ti^*$ | $L1_2$ | $L1_2$ | Y |
| Sc | | | | | | | | | $L1_2$ | $L1_2$ | $L1_2$ | $L1_2$ | Sc |
| Zr | | | $SV_3$ | $D0_a$ / $SV_3$ | | | | | $L1_2$ | $L1_2$ | $D0_{24}$ | $D0_{24}$ | Zr |
| Hf | | | | | | | | | $L1_2$ | $L1_2$ | $D0_{24}$ | $D0_{24}$ | Hf |
| Ti | $Mo_3Ti^*$ | $Mo_3Ti^*$ | $A15$ | | $A15$ | $A15$ | | | $L1_2$ | $L1_2$ | $D0_{24}$ / $PuAl_3$ | $D0_{24}$ | Ti |
| Nb | $A15$ | $L6_0$ | $A15$ | $A15$ | $A15$ | $Nb_3Pd^\dagger$ | $D0_{24}$ | $L1_2$ | $L1_2$ / $Co_3V$ | $Co_3V$ | $NbPt_3$ | $NbPd_3$ | Nb |
| Ta | $A15$ | $FCC^{[001]}_{AB3}$ | $A15$ | $A15$ | | | | | $L1_2$ / $Co_3V$ | $L1_2$ | $NbPt_3$ | $D0_{22}$ | Ta |
| V | $A15$ / $D0_3$ | $Mo_3Ti^*$ | $A15$ | $A15$ | $A15$ | $A15$ | $Re_3Ru^*$ | $Re_3Ru^*$ | $L1_2$ / $D0_{19}$ | $L1_2$ / $D0_{19}$ | $D0_{22}$ / $D0_a$ | $D0_{22}$ / $NbPd_3$ | V |
| Mo | $A15$ | | $A15$ | | $D0_{19}$ | | $D0_{19}$ | | $D0_{19}$ | $D0_{19}$ | ? | | Mo |
| W | | | | | | | $D0_{19}$ | $\sigma$ / $D0_{19}$ | $D0_{19}$ | $D0_{19}$ | | | W |
| Cr | $D0_{19}$ | | $A15$ | $A15$ | $L1_2$ | $L1_2$ | $A15$ | | $D0_{19}$ | $L1_2$ | $L1_2$ | | Cr |
| Tc | $D0_{19}$ | $D0_{19}$ | $D0_{19}$ | $D0_{19}$ | $D0_{19}$ | $D0_{19}$ | $D0_{19}$ | $D0_{19}$ | | | $FCC^{[001]}_{AB3}$ | | Tc |
| Re | $Re_3Ru^*$ | $Re_3Ru^*$ | $D0_{19}$ | $D0_{19}$ | $D0_{19}$ | $D0_{19}$ | $D0_{19}$ | $D0_{19}$ | | | ? / $FCC^{[001]}_{AB3}$ | | Re |
| Mn | | | $L1_2$ / $L6_0$ | $L1_2$ | $L1_2$ / $D0_{19}$ | | $D0_{19}$ | | $L1_2$ | $L1_2$ | $L1_2$ | $D0_{24}$ / $L1_2$ | Mn |
| Fe | | | $L6_0$ | $FCC^{[001]}_{AB3}$ | $L1_2$ | | | | $D0_{22}$ | $D0_{24}$ | $L1_2$ | $L1_2$ / $D0_{23}$ | Fe |
| Os | | $D0_a$ | | | | | | $D0_a$ | | | | | Os |
| Ru | $D0_a$ | | $L1_2$ | | | | $D0_a$ | | $D0_{19}$ | | | | Ru |
| Co | | | | | $D0_{19}$ | | | | | | $L1_2$ | $FCC^{[001]}_{AB3}$ | Co |
| Ir | | $L1_2$ | | $W3$ | | | | $D0_{19}$ | | | | | Ir |
| Rh | | | | | $D0_{22}$ | | | | $W3$ | | | | Rh |
| Ni | | | | ? / $D0_{22}$ | | | | | | | $D0_{23}$ | $FCC^{[001]}_{AB3}$ | Ni |
| Pt | | | | | $L1_2$ | | | | | $D0_{22}$ | | $L1_3$ | Pt |
| Pd | | | | $L1_3$ | | | | | | $L1_2$ | | | Pd |
| Au | | | | | | ? / $D0_{23}$ | | | | | | $L1_2$ | Au |
| Ag | | | | $L1_2$ | | $D0_{23}$ | | | | ? | | $L1_3$ | Ag |
| Cu | | | | | $L1_2$ | $L1_2$ / $D0_{23}$ | | | | ? / $L1_3$ | $L1_3$ | | Cu |
| Hg | | | | | | | | | | ? / $L1_2$ | $D0_{22}$ | | Hg |
| Cd | | | | ? | ? | | | | | $L1_2$ / $L1_3$ | $D0_{22}$ | | Cd |
| Zn | | $L1_2$ | $NbPd_3$ | $D0_{23}$ | $D0_{22}$ | $D0_{22}$ | | | | | $L1_2$ / $L1_3$ | | Zn |

FIG. 4: A Pettifor-type structure map for 1:3 stoichiometry compounds in PGM binary systems. The symbols are as in Fig. 2.

To help accelerate this process of experimental validation, discovery and development of materials [54] we are in the process of setting up a public domain REST-API that will allow the scientific community to download information from the www.aflowlib.org repository. It would ultimately enable researchers to generate alloy information remotely on their own personal computers. Extension of the database to nano-alloys and nano-sintered systems is planned within the size-pressure approximation (i.e. Fig.2 of Ref. [55]), to study trends of solubility and size-dependent disorder-order transitions and segregation in nano-catalysts [21, 55–57], and nano-crystals [58, 59].

A few of our predictions correspond to phases where the driving force for ordering is small (i.e., the formation enthalpy is small and it may be difficult to reach thermal equilibrium), however, it should be noted that some experimentally reported phases have similarly small formation enthalpies. Some of these predicted phases could be more easily realized as nano-structured phases, where the thermodynamics for their formation may be more favorable. Our results should serve as the foundation for finite temperature simulations to identify phases that are kinetically accessible. Rapid thermodynamical modelling and descriptor-based screening of systems predicted to harbor new phases should be used to pinpoint those with the greatest potential for applications [28]. Such simulations would be an invaluable extension to this work, however, the necessary tools to accomplish them on a similarly large scale are not yet mature.

### Acknowledgments

SC acknowledges support from DOD-ONR (N00014-13-1-0635, N00014-11-1-0136, N00014-09-1-0921). GLWH is grateful for support from the National Science Foundation, DMR-0908753. OL thanks the Center for Materials Genomics of Duke University for its hospitality. The authors thank Dr. K. Rasch and Dr. C. E. Calderon for useful comments.

# Appendix

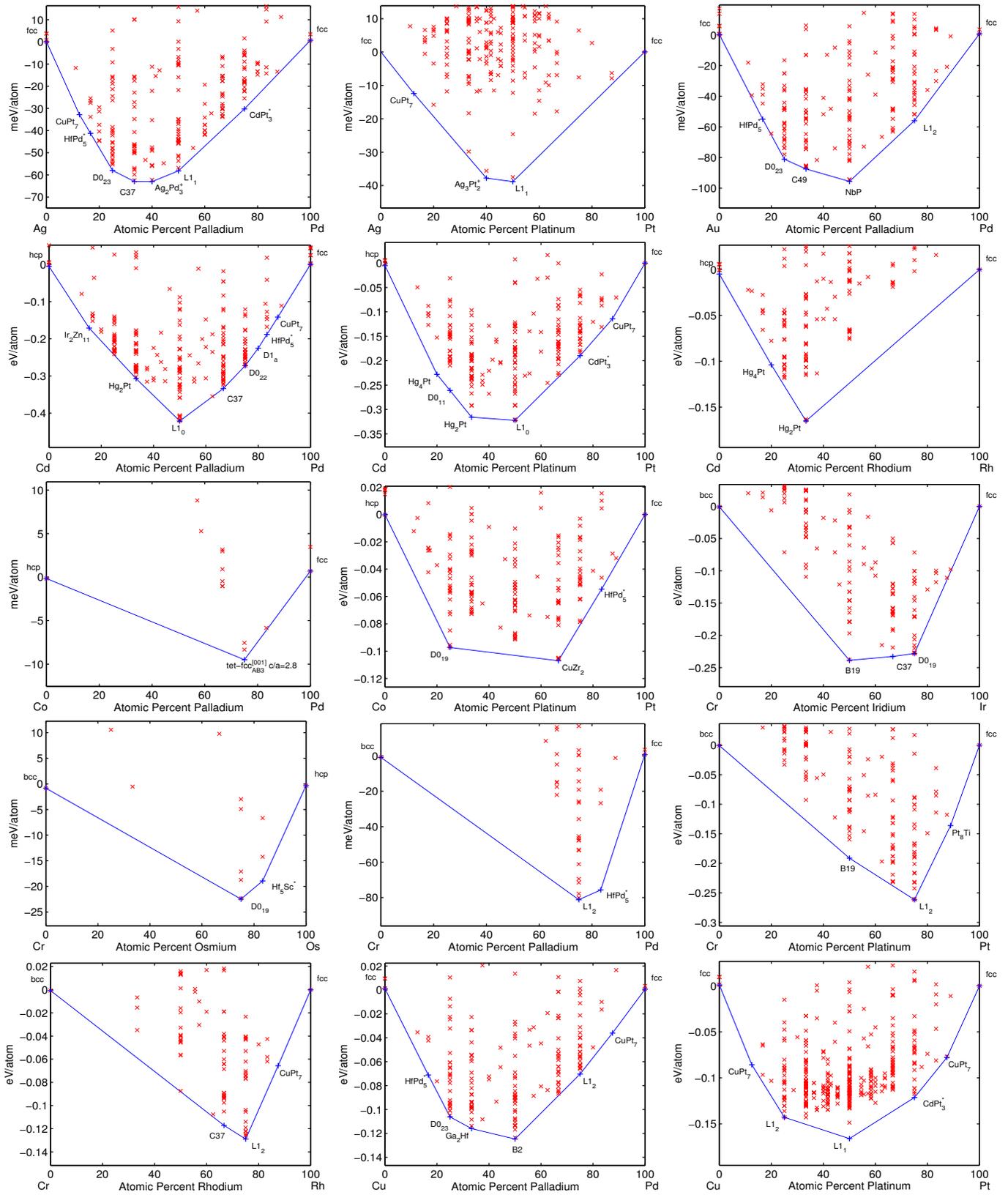

FIG. 5: Convex hulls for the systems AgPd, AgPt, AuPd, CdPd, CdPt, CdRh, CoPd, CoPt, CrIr, CrOs, CrPd, CrPt, CrRh, CuPd, and CuPt.

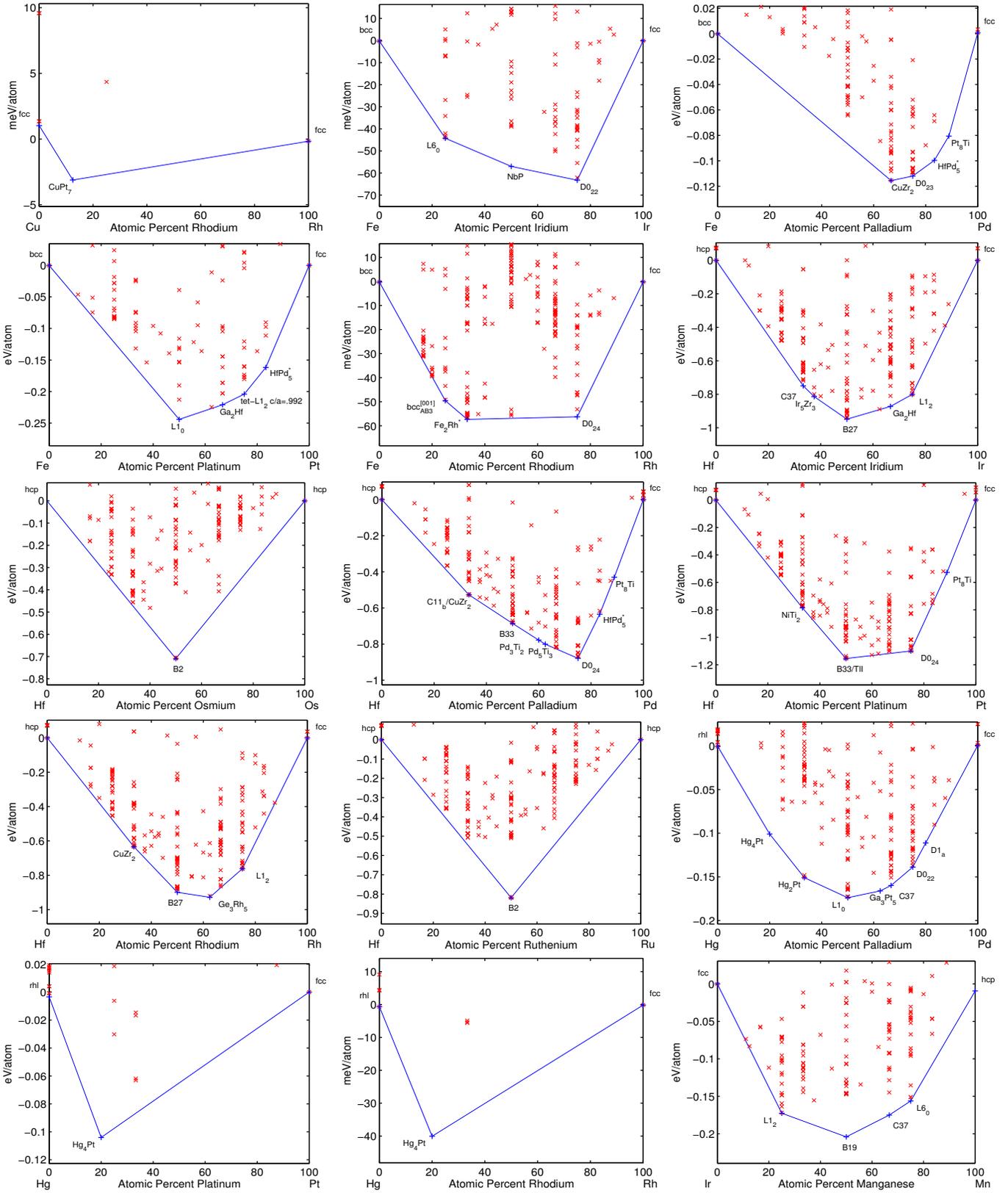

FIG. 6: Convex hulls for the systems CuRh, FeIr, FePd, FePt, FeRh, HfIr, HfOs, HfPd, HfPt, HfRh, HfRu, HgPd, HgPt, HgRh, and IrMn.

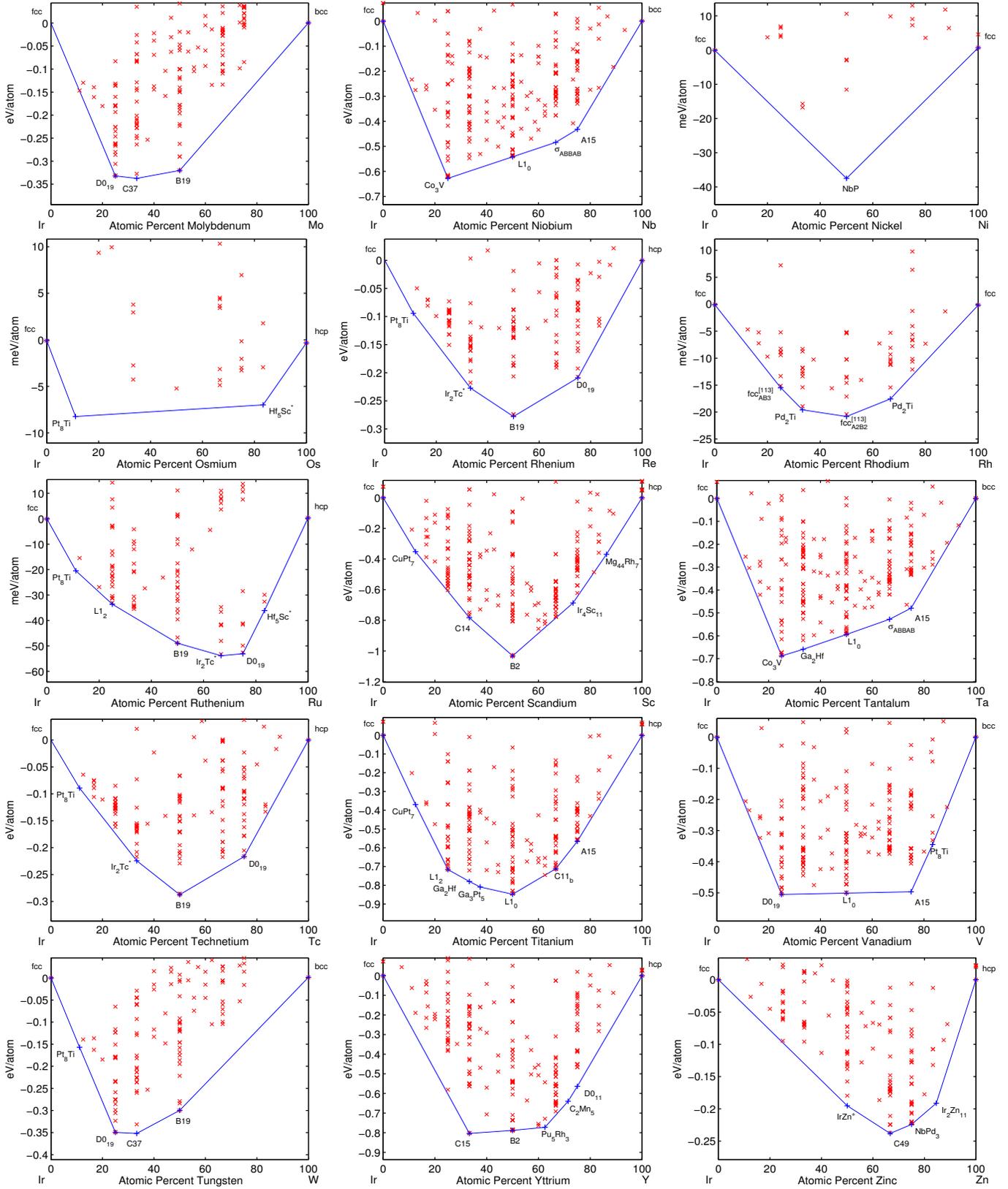

FIG. 7: Convex hulls for the systems IrMo, IrNb, IrNi, IrOs, IrRe, IrRh, IrRu, IrSc, IrTa, IrTc, IrTi, IrV, IrW, IrY, and IrZn.

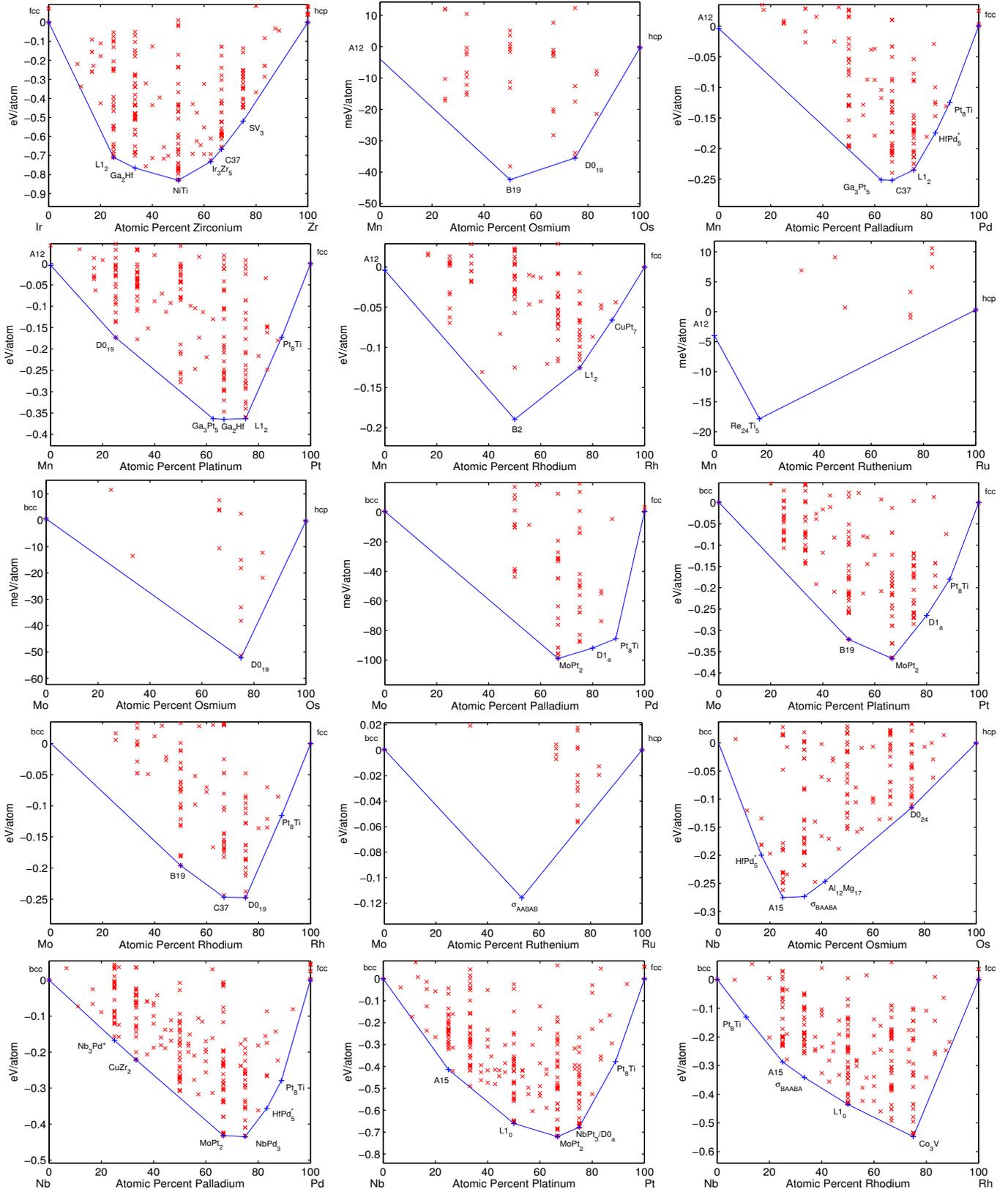

FIG. 8: Convex hulls for the systems IrZr, MnOs, MnPd, MnPt, MnRh, MnRu, MoOs, MoPd, MoPt, MoRh, MoRu, NbOs, NbPd, NbPt, and NbRh.

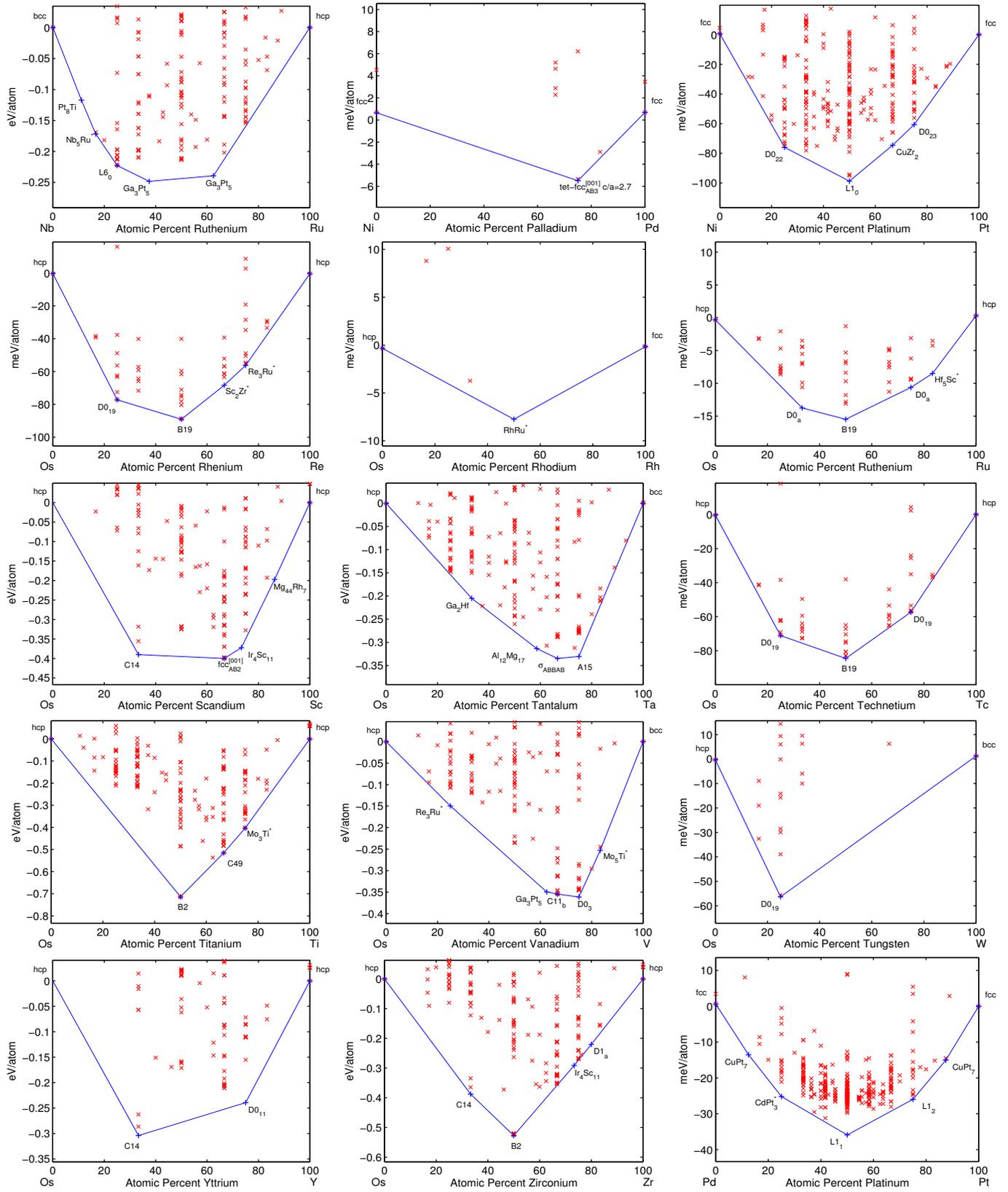

FIG. 9: Convex hulls for the systems NbRu, NiPd, NiPt, OsRe, OsRh, OsRu, OsSc, OsTa, OsTc, OsTi, OsV, OsW, OsY, OsZr, PdPt, and PdRe.

FIG. 10: Convex hulls for the systems PdSc, PdTa, PdTc, PdTi, PdV, PdW, PdY, PdZn, PdZr, PtRe, PtRh, PtRu, PtSc, PtTa, and PtTc.

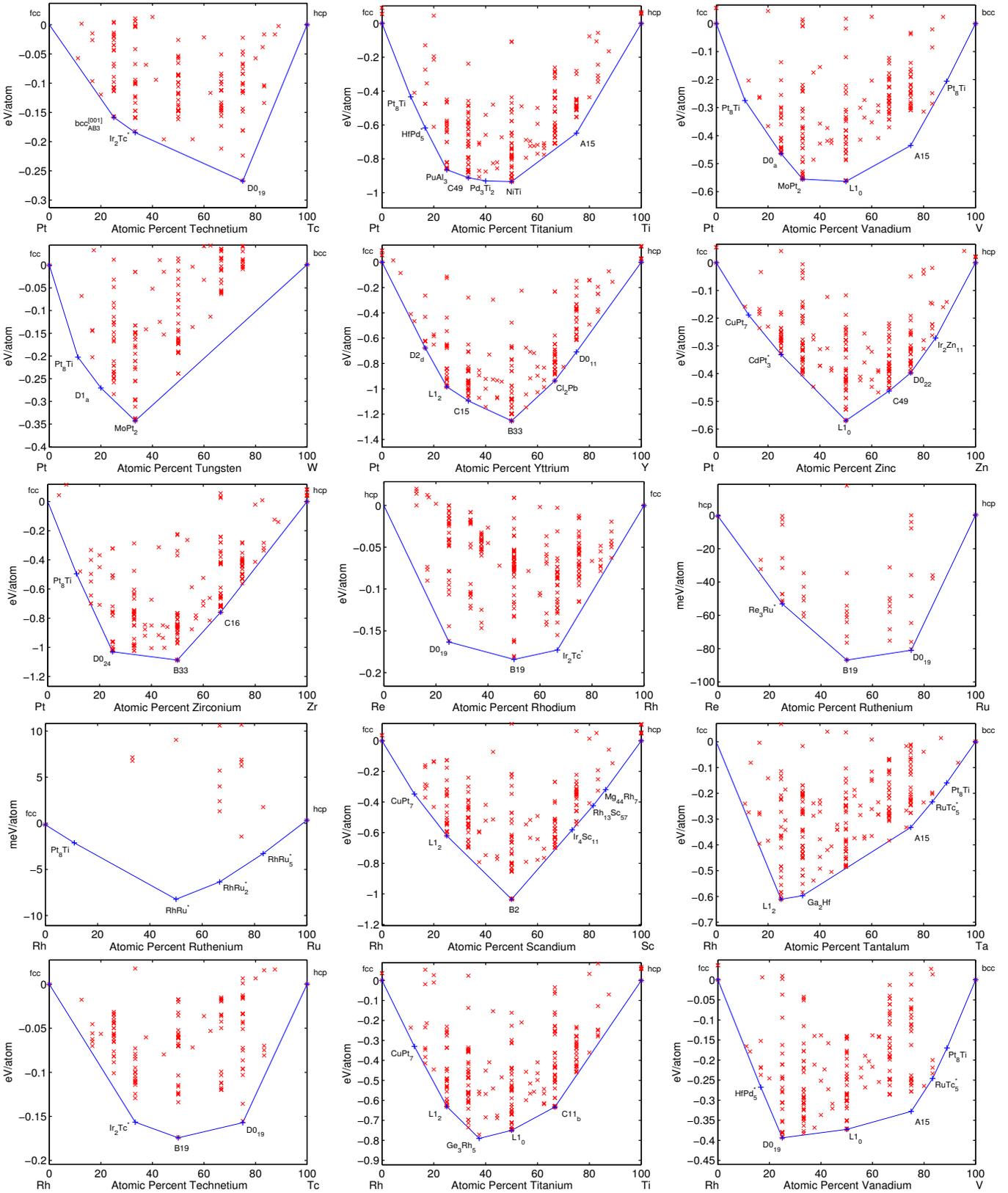

FIG. 11: Convex hulls for the systems PtTi, PtV, PtW, PtY, PtZn, PtZr, ReRh, ReRu, RhRu, RhSc, RhTa, RhTc, RhTi, RhV, and RhW.

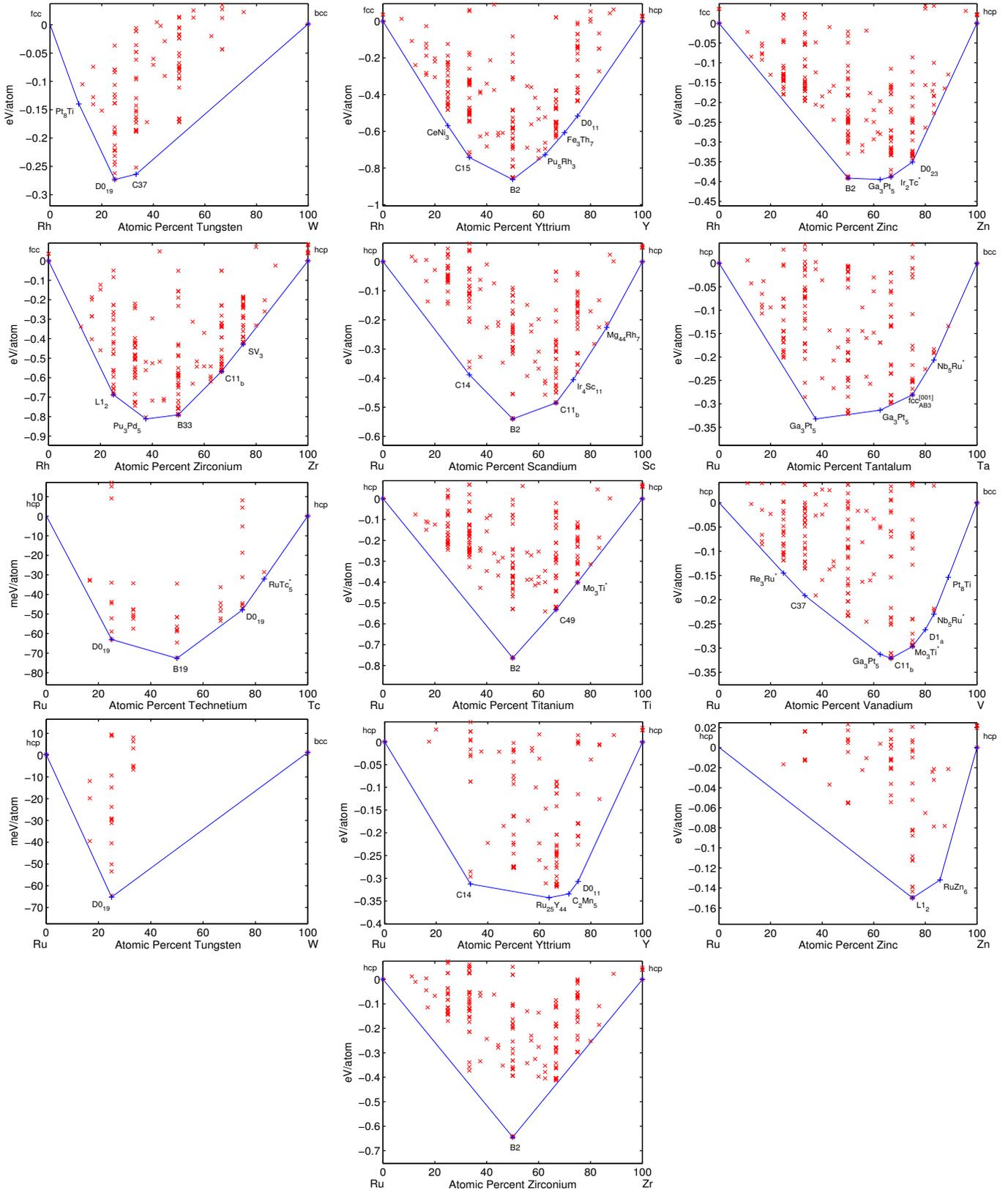

FIG. 12: Convex hulls for the systems RhY, RhZn, RhZr, RuSc, RuTa, RuTc, RuTi, RuV, RuW, RuY, RuZn, and RuZr.